\documentclass[twocolumn, aps, superscriptaddress]{revtex4-1}

\usepackage{amsfonts}
\usepackage{amsmath}
\usepackage{graphicx}
\usepackage{color}

\def\ket#1{\mathinner{|{#1}\rangle}}

\begin{document}

\title{Poking holes and cutting corners to achieve Clifford gates with the surface code}

\author{Benjamin J. Brown}
\email{benjamin.brown@nbi.ku.dk}
\affiliation{Niels Bohr International Academy, Niels Bohr Institute, Blegdamsvej 17, 2100 Copenhagen, Denmark}
\author{Katharina Laubscher}
\affiliation{Department of Physics, University of Basel, Klingelbergstrasse 82, CH-4056 Basel, Switzerland}
\author{Markus S. Kesselring}
\affiliation{Department of Physics, University of Basel, Klingelbergstrasse 82, CH-4056 Basel, Switzerland}
\affiliation{Dahlem Center for Complex Quantum Systems, Freie Universit\"{a}t Berlin, 14195 Berlin, Germany}
\author{James R. Wootton}
\affiliation{Department of Physics, University of Basel, Klingelbergstrasse 82, CH-4056 Basel, Switzerland}

\date{\today}

\begin{abstract}
The surface code is currently the leading proposal to achieve fault-tolerant quantum computation. Among its strengths are the plethora of known ways in which fault-tolerant Clifford operations can be performed, namely, by deforming the topology of the surface, by the fusion and splitting of codes and even by braiding engineered Majorana modes using twist defects. Here we present a unified framework to describe these methods, which can be used to better compare different schemes, and to facilitate the design of hybrid schemes. Our unification includes the identification of twist defects with the corners of the planar code. This identification enables us to perform single-qubit Clifford gates by exchanging the corners of the planar code via code deformation. We analyse ways in which different schemes can be combined, and propose a new logical encoding. We also show how all of the Clifford gates can be implemented with the planar code without loss of distance using code deformations, thus offering an attractive alternative to ancilla-mediated schemes to complete the Clifford group with lattice surgery.
\end{abstract}

 \maketitle

\section{Introduction}
\label{sec:intro}

The components that make up quantum technologies are inherently sensitive to noise. This is a problem which, if unresolved, will inhibit the scalability of quantum information processing tasks. To overcome this issue quantum error-correcting codes have been developed~\cite{Kitaev03, Dennis02, LidarBrun, Terhal15, Brown16a}, where logical quantum states are redundantly encoded in robust subspaces of the Hilbert space of many physical qubits. As such, there is a considerable number of physical qubits of a scalable quantum computer that are dedicated to the role of error correction.

Given the prohibitive cost of quantum resources, it is important to discover fault-tolerant schemes for universal quantum computation that use as few physical qubits as possible. Among other factors, the resource cost of fault-tolerant quantum computation depends on the choice of quantum error-correcting codes into which we choose to encode quantum information, and the different schemes we use to implement a universal set of computational gates. Indeed, there has been considerable effort dedicated to minimising the resource cost of quantum computation.

A leading approach to realising fault-tolerant quantum computation is broadly known as topological quantum computation~\cite{Kitaev03, PreskillsLecture, Nayak08, Pachos}. With this approach, we protect quantum information by encoding it into non-local degrees of freedom, using objects including non-Abelian anyons~\cite{Kitaev03, Kitaev06}, punctures~\cite{Raussendorf06, Raussendorf07, Raussendorf07a, Bombin08, Bombin09, Fowler09, Fowler12a, Brell15, Cong16, Cong17}, or by use of extrinsic defects~\cite{Bombin10, Bombin11, BarkeshliJianQi13, Barkeshli14, Hastings15, Wootton15a}, otherwise known as twists. With these schemes, encoded information is manipulated by braiding these objects to realize fault-tolerant universal quantum computation. In addition to schemes where protected quantum information undergoes unitary rotations by braiding, other promising schemes are known where quantum information is encoded over non-contractable cycles of a lattice of physical qubits which are embedded on a manifold with non-trivial topology~\cite{Kitaev03, Dennis02}. Fault-tolerant entangling gates are then achieved either transversally~\cite{Bombin06, Bombin07a, Bombin15}, or by lattice surgery~\cite{Horsman12, Landahl14}.

It is the goal of this manuscript to unify some of these schemes by consideration of a specific lattice model. Here we unify two low-overhead approaches to encoding qubits using the surface code~\cite{Kitaev03, Bravyi98, Dennis02}, namely, lattice surgery~\cite{Horsman12, Landahl14}, together with defect encoding schemes~\cite{Bombin10, Hastings15}. This unification is made using a correspondence between the corners of the planar code~\cite{Dennis02} and twist defects~\cite{Bombin10}.

While it is well known in generality that non-Abelian defects exist on system boundaries in between distinct phases in the topological condensed-matter literature~\cite{Beigi11, Kong12, Lindner12, Levin13, Barkeshli13, Barkeshli14}, here we find it instructive to consider boundaries of a very specific lattice system. In particular, we use the correspondence between twist defects and Majorana fermions~\cite{Bombin10, Brown13, Wootton15a, Zheng15} to realize logical gates by braiding twist defects. Specifically, we will show that we can achieve a fault-tolerant realisation of the full Clifford group in a two-dimensional system using lattice surgery methods, and by braiding twist defects via code deformations.

Further to this, we also consider interactions between different topological schemes for encoding quantum information. In particular, we consider how punctures interact with twist defects, and show that we can use holes to perform fault-tolerant measurement-only topological quantum computation~\cite{Bonderson08, Bonderson09, Hastings15} with twist defects, by braiding twists and holes. We also show that we can encode logical qubits in hybridized schemes that use both holes and twist defects to encode logical qubits. Such encodings are of interest as they may enable new schemes that realize fault-tolerant quantum computation with lower overhead demands. See, for instance, related recent work~\cite{Yoder16} following this goal where logical qubits are encoded on a surface code with three corners and a single central twist.

The paper is organized as follows; in Sec.~\ref{sec:Review} we give a brief overview of some of the different approaches that have been followed to achieve low-overhead fault-tolerant quantum computation. Then, in Sec.~\ref{sec:prelim} we introduce notation and review the stabilizer formalism, anyon models, the Clifford group, and different methods of encoding qubits using the surface code. We also develop a diagrammatic language that we use to build correspondences between different schemes for encoding qubits. In Sec.~\ref{Sctn:SingleQubitGates} we review code deformation, and show that we can braid the corners of the planar code using code deformation to perform Clifford gates. In Sec.~\ref{Sctn:EntanglingGates} we consider interactions between twists and holes, and show that we can use holes to perform fault-tolerant parity measurements between qubits encoded using twist defects in a measurement-only topological quantum computational scheme. In Sec.~\ref{Sctn:Surgery} we illustrate a connection between lattice surgery and measurement-only topological quantum computation. Finally, in Sec.~\ref{Sctn:Hybrids} we introduce a new hybrid scheme of encoding qubits and discuss its advantages and drawbacks before giving some concluding remarks in Sec.~\ref{Sctn:Conclusion}. We point out that further details and alternative explanations of some aspects of this manuscript can be found in the masters project work on which it is based, see Ref.~\cite{Laubscher16}.

\section{Fault-tolerant topological quantum computation schemes}
\label{sec:Review}
This work seeks to develop and unify tools for surface code quantum computation. Before we present our results, we first comment on other promising topological quantum computational schemes. Ultimately, the resource cost of a computational scheme will depend on how the logical error rate of a code scales as a function of the number of physical qubits used, together with time and space requirements that are needed to execute a universal gate set with the chosen system. Certainly, it is important to consider all of these factors when trying to determine the number of physical qubits a particular architecture will need.

Given a system that can perform Clifford gates and prepare noisy copies of magic states, universal quantum computation can be achieved via magic state distillation~\cite{BravyiKitaev05}. A recent review of developments in magic state distillation protocols can be found in Ref.~\cite{Campbell16b}. As such, we first restrict our attention to achieving Clifford gates in two-dimensional architectures.

Previously, using the surface code where qubits are encoded using punctures, the full Clifford group is completed using ancillas that are prepared in the Y-state~\cite{Fowler12a}, i.e., an eigenstate of the Pauli-Y matrix. Logical qubits are prepared in the Y-state via a probabilistic and noisy process~\cite{Fowler12c}. Once prepared, the distilled states can be used to perform an arbitrary number of phase gates~\cite{Aliferis07, Jones12}. Nevertheless, the requirement for nearby ancilla qubits prepared in the Y-states, together with the initial overhead cost of preparing these Y-states, will contribute to the resource cost of quantum computation compared with schemes where phase gates can be achieved natively. Alternatively, by addition of lattice dislocations~\cite{Hastings15}, the Clifford group can be achieved by making parity measurements between logical qubits. Indeed, given arbitrary two-qubit Pauli parity measurements, arbitrary two-qubit Clifford gates can be achieved~\cite{Terhal15}. The ability to perform arbitrary two-qubit gates circumvents the need for single-qubit Clifford gates to generate the Clifford group. Notably, the dislocation code scheme involves introducing a small number of weight-five stabilizer measurements to the surface code. 

The two-dimensional color code also achieves the full Clifford group logical gates via transversal operations~\cite{Bombin06}. While transversal operations are very appealing compared with the code deformation schemes that seem to be required for quantum computation with the surface code, realising the color code comes at the expense of increased weight stabilizer measurements. Specifically, the surface code requires weight-four stabilizer measurements compared with weight-six measurements that are required of the color code.

It is also worth mentioning the gauge color code~\cite{Bombin15} which, notably, performs a universal transversal gate set via gauge fixing~\cite{Paetznick13, Anderson14}, and may thus offer a reduction in overhead compared with magic-state distillation based schemes of computation. While this is an appealing feature, the gauge color code is three dimensional, and as such, is challenging to realize with locally interacting qubits arranged on a two-dimensional surface. While some effort has been made to reduce the engineering demands of realising three-dimensional codes via dimension jumping~\cite{Bombin15b}, or by finding two-dimensional variants of the gauge color code~\cite{Bravyi15, OConnor16, Jones16}, these schemes still, respectively, require either some three-dimensional components, or come at the expense of the threshold error rate.

While gauge fixing with the gauge color code offers an elegant approach to achieving a universal gate set, the space-time quantum resource cost scales equally~\cite{Brown16, Campbell16, Campbell16a} with other proposals, up to a constant factor, by use of single-shot error correction~\cite{Bombin15a, Brown16}. As such, despite the apparent advantages of these codes, current proposals for magic-state distillation based schemes with two-dimensional architectures remain attractive due to their practicality and high error thresholds~\cite{Raussendorf06}.

\section{Encoding qubits with the surface code}
\label{sec:prelim}

In this Section we review the backgorund material we use throughout the present paper including stabilizer formalism~\cite{GottesmanThesis}, anyon models~\cite{Kitaev03, PreskillsLecture, Kitaev06, Nayak08, Pachos}, the Clifford group, and different methods of encoding qubits with the surface code~\cite{Dennis02}.

\subsection{The stabilizer formalism}

Quantum states are robustly maintained in the code space of a quantum error-correcting code. We specify the code space of a stabilizer quantum error-correcting code with its stabilizer group, $\mathcal{S}$. The stabilizer group is an Abelian subgroup of the Pauli group, $\mathcal{P}$, with $ - \openone \not\in \mathcal{S}$. Up to phases, the Pauli group is generated by the standard Pauli matrices, $X_j$, $Y_j$, and $Z_j$, where $j$ indicates the qubit of the system the operator acts on.

The code space of a stabilizer code is the common $+1$ eigenspace of all the elements of the stabilizer group $ s \in \mathcal{S}$, i.e.
\begin{equation}
s \ket{\psi} = (+1) \ket{\psi},\quad \forall s, \label{Eqn:Stabilizer}
\end{equation}
where $\ket{\psi}$ are basis vectors that span the code space. Stabilized states, $\ket{\psi}$, are commonly known as codewords.

We act on the code space with logical operators, $\overline{X}_k,\, \overline{Z}_k$, which are distinguished from other operators with bar notation. Logical operators commute with all members of the stabilizer group, but are not themselves members of the stabilizer group. The logical operators generate the logical Pauli group which acts on the code space. They therefore satisfy the following properties; $\overline{X}_k \overline{Z}_k = - \overline{Z}_k \overline{X}_k $, and $\overline{X}_k \overline{Z}_l=  \overline{Z}_l \overline{X}_k$ for $k \not= l$.

An important quantity to introduce that characterises stabilizer codes is the code distance, which is commonly denoted $d$. The code distance is the smallest set of qubits which support one non-trivial logical operator of the code, where the support of an operator are the set of qubits an operator acts upon non-trivially. As a first order approximation, the code distance quantifies the ability of a code to tolerate noise, as it denotes the smallest number of qubits that must be rotated in order to complete a logical operation on the code subspace of the code.

It will also be helpful to note that the action of a logical operator on the code space is invariant if the logical operator is multiplied by an element of the stabilizer group. Specifically, two logical operators $\overline{L}$ and $\overline{L}' = s \overline{L} $ for $s \in \mathcal{S}$ satisfy relationship $ \overline{L}' | \psi \rangle = \overline{L} | \psi \rangle $ for all codestates $| \psi \rangle$. This follows from Eqn.~(\ref{Eqn:Stabilizer}), and the commutation relation of elements of the stabilizer group with the logical operators. This will be useful as it allows us to change, or `clean'~\cite{Bravyi09}, the support of logical operators, such that certain qubits of the stabilizer code do not support certain choices of logical operator.

\subsection{Anyons}
\label{subsec:anyons}

A complementary and natural way to understand schemes of topological quantum error correction is through the language of anyonic excitations. Anyons are point-like quasiparticles whose motion is restricted to two spatial dimensions. This restriction allows exotic exchange behaviour to arise.

We will frequently invoke this quasiparticle picture to elucidate the physics of the error-correcting codes we study. For a detailed description of anyon models, we refer the reader to Refs.~\cite{Kitaev03, PreskillsLecture, Nayak08, Pachos} and Appendix~E of Ref.~\cite{Kitaev06}. Here we briefly review two explicit anyon models that will be relevant throughout this Manuscript.

\subsubsection{The $D(\mathbb{Z}_2)$ anyon model}
\label{Subsubsctn:Z2}

The anyon model of the surface code is known as $D(\mathbb{Z}_2)$~\cite{Kitaev03}. It is composed of four anyons, $e$, $m$, and, $\psi$, together with the vacuum particle, $1$, which denotes `no particle'. All anyon models include the vacuum particle.

For historical reasons, the $e$ and $m$ anyons are known as the electric and magnetic charges. These anyons are their own antiparticles. This means that they will annihilate if combined.
The combination of pairs of particles is captured by the notion of `fusion', denoted by the binary operation `$\times$'. The fact that $e$ and $m$ excitations are their own antiparticles are captured by the fusion rules
$$
e \times e = m \times m = 1.
$$

Exchanging pairs of $e$ quasiparticles gives rise to bosonic exchange statistics, in the sense that they realize a trivial, i.e. $+1$, phase upon the exchange. The same is true when pairs of $m$ particles are exchanged. Braiding an $e$ and an $m$ excitation however has a non-trivial effect. This is seen when one particle is moved through a full loop around the other, which is known as a monodromy. Braiding $e$ around an $m$, or vice versa, introduces a global phase of $-1$ to the system.

The $\psi$ anyon is a particle that is composed of an $e$ and an $m$ excitation, which is specified by the fusion rule
$$
e \times m  = \psi.
$$

The particle labeled $\psi$ has fermionic exchange behaviour, and so their wavefunction acquires $-1$ global phase upon exchange. This arises from the non-trivial exchange behaviour of the component $e$ and $m$ anyons.

\subsubsection{The Ising anyon model}
\label{Subsubsctn:Ising}
The Ising anyon model~\cite{Moore91, Kitaev06} has two non-trivial anyon types, $\sigma$ and $\psi$. The particle $\psi$, as above, is a fermion that is its own antiparticle. The equivalence between these particles will be used in this work, and so any reference to fermions refers interchangeably to both.

The $\sigma$ anyon is non-Abelian. Fusing two Ising anyons can result in either annihilation where the vacuum particle is produced, or the fusion outcome can be one fermion. This is captured by the fusion rule
$$
\sigma \times \sigma = 1 + \psi.
$$

The $\sigma$ particle is also able to absorb a fermion, which is represented by the fusion rule
$$
\sigma \times \psi = \sigma.
$$
The number of fermions absorbed by a pair of Ising anyons can be learned by fusing the pair.

The $\sigma$ anyons are equivalent to Majorana modes, and can be well described by Majorana operators~\cite{Kitaev06,Wootton15a}. We will not require this description in this work. However, we note one important feature. This is that the Majorana parity operator, which assigns a phase of $\pm1$ depending on whether a pair of $\sigma$ anyons will fuse to vacuum or a $\psi$ particle, is equivalent, up to a global phase, to the operator which fuses a fermion with each $\sigma$ particle. These operators are also equivalent to a full monodromy of one $\sigma$ particle around the other. It is these operators that are used as the logical Pauli operators of qubits encoded with Ising anyons. A single exchange of two Ising anyons therefore implements a unitary rotation that corresponds to the square root of Pauli operator. Such rotations are members of the Clifford group, which we next discuss.

\subsection{The Clifford group}

Elements of the Clifford group, $U \in \mathcal{C}$, map elements of the Pauli group onto elements of the Pauli group under conjugation. It is defined
\begin{equation}
\mathcal{C} = \left\{ U : \forall P \in \mathcal{P} , \quad  U P U^\dagger \in \mathcal{P} \right\}.
\end{equation}
The Clifford group can be generated by two single-qubit unitary rotations, or `gates', the phase gate and the Hadamard gate, which, respectively, can be expressed in terms of Pauli matrices such that
\begin{equation} 
S = (e^{\text{i} \pi /4} \openone + e^{-\text{i}\pi /4} Z ) / \sqrt{2}, 
\quad
H = (X + Z) / \sqrt{2},
\end{equation} 
together with a two-qubit controlled-not gate, which is also Hermitian
\begin{equation} 
\text{CNOT} = ( \openone \otimes \openone + Z \otimes \openone   + \openone \otimes X - Z \otimes X) / 2.
\end{equation}
The Clifford group act on the Pauli matrices as follows. The phase gate, or `S-gate', obeys the following equations
\begin{equation}
S X  S^\dagger = -Y,  \quad S Y  S^\dagger = X,\quad  SZS^\dagger = Z,
\end{equation}
the Hadamard gate, which is Hermitian, satisfies
\begin{equation}
H X  H = Z,  \quad H YH = -Y, \quad  HZH = X,
\end{equation}
and for the controlled-not gate we have
$$
\text{CNOT} \, (X \otimes \openone) \, \text{CNOT} = X \otimes X,
$$
$$
\text{CNOT}\,  (\openone \otimes X) \, \text{CNOT} = \openone \otimes X, 
$$
$$
\text{CNOT}\,  (Z \otimes \openone) \,  \text{CNOT} = Z \otimes \openone, 
$$
\begin{equation}
\text{CNOT} \, (\openone \otimes Z) \, \text{CNOT} = Z \otimes Z. \label{eqn:CNOT}
\end{equation}

\subsection{The planar code}
\label{subsec:surfcode}
\begin{figure}
\includegraphics{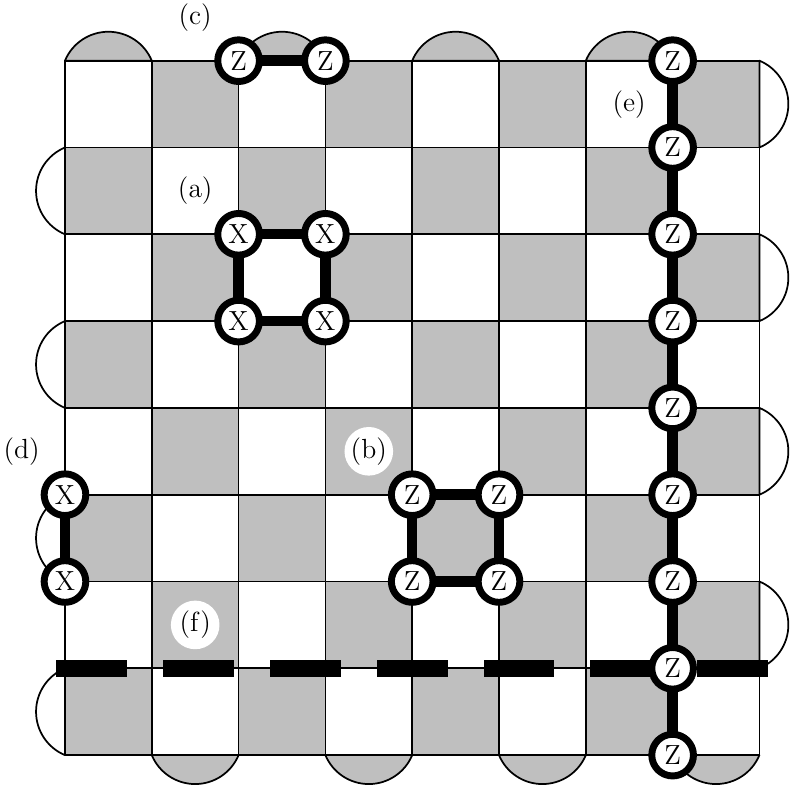}
\caption{The planar variant of the surface code with two rough boundaries and two smooth boundaries. Qubits lie on the vertices of the square lattice. (a)~and (b)~show the two different types of weight-four stabilizer operators. (c)~and (d)~show two stabilizers that lie at the boundary of the lattice. (e)~The logical operator $\overline{Z}$ is the tensor product of Pauli-Z operators extending from the top to the bottom of the lattice. (f)~The logical operator $\overline{X}$ is the tensor product of Pauli-X operators along the horizontal dashed line that extends from the left to the right of the lattice.}
\label{fig:WPM}
\end{figure}

The planar code~\cite{Bravyi98, Dennis02} is defined on an $L \times L$ square lattice with one qubit placed on each vertex of the lattice, as shown in Fig.~\ref{fig:WPM}. The distance of the code is $d = L$. This representation of the planar code is given in Ref.~\cite{Wen03}, but it is easily checked~\cite{Nussinov09, Brown11} that this representation, up to local unitary operations, is equivalent to the more conventional representation of the surface code~\cite{Kitaev03, Dennis02} where the qubits lie on the edges of a square lattice.

To specify the stabilizer group we bicolour the faces, $f$, of the lattice black and white as in Fig.~\ref{fig:WPM}. With this colouring we can write the two different types of stabilizer that form the stabilizer group, namely, we have the operator 
\begin{equation}
A_f = \prod_{j \in \partial f} X_j, \label{Eqn:Star}
\end{equation}
on each white face of the lattice, and the operator
\begin{equation}
B_f = \prod_{j \in \partial f} Z_j, \label{Eqn:Plaq}
\end{equation}
on each black face of the lattice. The set $\partial f$ denotes the qubits that touch face $f$. We show an example of an $A_f$ and a $B_f$ operator in Figs.~\ref{fig:WPM}(a) and~\ref{fig:WPM}(b), respectively.

We must also define stabilizers on the boundary of the planar code. The boundary stabilizers are shown in Fig.~\ref{fig:WPM} by adding additional faces to the boundary of the lattice. We have added black faces to the top and the bottom of the lattice, and white faces to the left and right sides of the lattice.  With the addition of these extra faces, the boundary terms are specified with the definitions given in Eqns.~(\ref{Eqn:Star}) and~(\ref{Eqn:Plaq}). We show explicit examples of boundary terms in Figs.~\ref{fig:WPM}(c) and~(d).

Provided the stabilizer group contains only commuting generators, we have a lot of freedom as to which types of faces we may like to add to the boundary. Indeed, the choice of boundary stabilizers plays a very important role on the encoding properties of the planar code, see~\cite{Bravyi98}. To maintain consistency with the terminology used in Refs.~\cite{Bravyi98, Dennis02}, we call boundaries with black faces on the boundary `rough boundaries' and boundaries with white faces on the boundary `smooth boundaries'. In Fig.~\ref{fig:WPM} we have rough boundaries on the top and bottom of the lattice, and smooth boundaries on the left and right sides of the lattice.

With the choice of boundaries shown in Fig.~\ref{fig:WPM}, the planar code encodes one logical qubit. The logical operators are strings of Pauli operators that extend between different boundaries of the same type. The logical operator $\overline{Z}$ is the tensor product of Pauli-Z operators that are supported along a string that extends from the top to the bottom of the lattice, between two rough boundaries. We show $\overline{Z}$ in Fig.~\ref{fig:WPM}(e). The logical operator $\overline{X}$ is the tensor product of Pauli-X operators supported on a string that runs from the left side to the right-hand side of the lattice. The qubits that support $\overline{X}$ lie on the thick dashed line shown in Fig.~\ref{fig:WPM}(f). Unlike $\overline{Z}$, the operator $\overline{X}$ is a string that stretches between two different smooth boundaries of the lattice. It is easily seen that $\overline{X}$ anti commutes with $\overline{Z}$, as they are commonly supported on only one qubit of the lattice.

It is worth noting that, provided the string operators commute with the stabilizer group, we are free to choose logical operators along any string on the lattice as long as the terminal boundaries of the string are not changed. This is easily seen by observing that we can manipulate the path of the strings of logical operators shown in Fig.~\ref{fig:WPM} by multiplying them by elements of the stabilizer group. We frequently make use of this fact in discussions in later Sections of the present Manuscript.

The logical operators of the planar code can be understood using the picture of anyonic quasi-particle excitations. Indeed, as it is explained in~\cite{Kitaev03}, string-like Pauli operators acting on the codewords of the planar code can be regarded as creation operators for anyons.
Specifically, a string operator of Pauli-Z(Pauli-X) operators with end points that lie in the bulk of the lattice create pairs of electric (magnetic) charges at the end points of the strings, as we have defined in Subsubsec.~\ref{Subsubsctn:Z2}. Hopping operators for $e$ and $m$ excitations also correspond to string-like operators.

Unlike the creation and hopping operators we have just mentioned, the string-like logical operators of the planar code do not create quasi-particle excitations. This is because the string operators terminate at boundaries of the planar code where quasi-particle excitations are absorbed, or `condensed'. Specifically, a rough(smooth) boundary is capable of absorbing $e$($m$) particles. We can therefore regard the $\overline{Z}$($\overline{X}$) logical operator as the process of creating a single $e$($m$) excitation at one rough(smooth) boundary, and subsequently transporting the particle across the lattice where it is then absorbed by the opposite rough(smooth) boundary. The braiding statistics of the exchange of these two excitations assure the appropriate commutation relations between these logical operators. We will frequently draw on this picture to demonstrate different logical operations using the surface code. Our freedom to deform logical operator strings is also elucidated in this picture, as we have argued that it is the role of logical operators to transport excitations between different boundaries. For the case of the planar code, the action of the operator is independent of its path along the lattice provided charges are transported between the appropriate boundaries. In general, the action of logical string operators are invariant under continuous deformations to their path over the lattice.

\subsection{Encoding logical qubits using holes}

\begin{figure}
\includegraphics{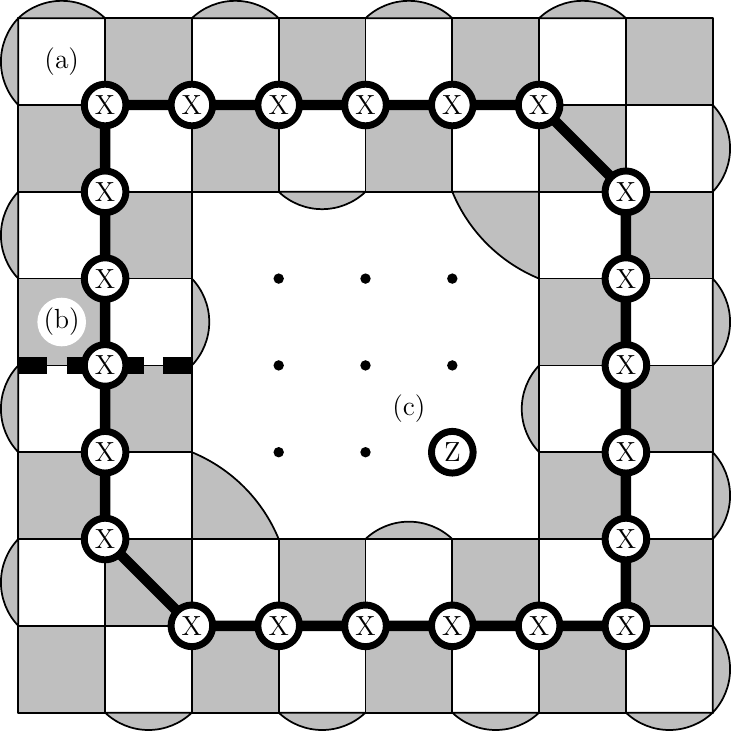}
\caption{A hole in the planar code modifies its code space. A hole is shown in the centre of the lattice. (a) When a hole is prepared, a new logical operator is produced which consists of a non-trivial cycle of Pauli-X operators that enclose the hole. The qubits that support the logical operator are marked by a thick dashed line. (b) The logical operator shown by label~(a) anti commutes with the conjugate logical operator which is composed of the tensor product of Pauli-Z operators that extends from the boundary of the hole to the boundary of the surface. (c) A hole can be produced in the lattice by measuring the qubits in the computational, i.e., Pauli-Z basis. \label{fig:Hole}}
\end{figure}

We can increase the number of encoded qubits for the surface code by introducing punctures to the lattice. We show a puncture, or hole, in Fig.~\ref{fig:Hole}. The qubits inside the puncture, shown as small black points, have been disentangled from the lattice. We point out that unlike the lattice shown in Fig.~\ref{fig:WPM}, the punctured lattice has only rough boundaries. As such, without a hole in the centre, this lattice would encode no logical qubits. By introducing a puncture in the centre of the lattice, we are able to encode a single logical qubit on the lattice. We show the logical operators for the encoded qubit in Figs.~\ref{fig:Hole}(a) and~(b). The logical operator $\overline{X}$ is the tensor product of Pauli-X operators supported on a cycle of qubits that enclose the puncture shown by a dashed line in the Figure. The logical operator $\overline{Z}$ is the tensor product of Pauli-Z operators supported along a line that connects the lattice boundary to the boundary of the puncture. We show such an operator in Fig.~\ref{fig:Hole}(b). As with the planar code, the logical operators here are string-like and allow continous deformations.

Like the planar code, the logical operators of the punctured planar code can also be interpreted from the point of view of topological excitations. Given that both the lattice boundary and the hole in the lattice have rough boundaries, they are both capable of absorbing an $e$~particle. Therefore the logical state of the hole relates to the number of electric charges that have been passed from the lattice boundary into the hole. Specifically, the parity of this number gives one two-level logical degree of freedom. The even case corresponds to vacuum due to the fusion rule $e\times e=1$, and the odd case correspond to an $e$ charge absorbed by the hole. We measure the charge parity absorbed by the hole with a measurement that encloses the hole, which is able to count the number of pairs of electric excitations that have been shared between the lattice boundary and the hole, thus giving the topological structure of the $\overline{X}$ logical operator. Indeed, this measurement is analogous to Gauss' law~\cite{Kitaev03}.

The distance of a code where qubits are encoded using holes depends on two quantities, the length of the boundary of a given hole, and the separation of a hole from another lattice boundary. As such, to maintain a code distance $d$, all holes with a common boundary type must be separated by a distance of at least $d$, and must maintain a distance of $d$ from the boundary of the lattice. Holes must also have a boundary of at least length $d$. As such, for the case of spherical holes, i.e. holes where the removed qubits are taken from a simply connected region, the radius of the region must be $\sim d$. For simplicity, we will only consider holes that cover simply connected regions.

In general, we can consider punctures with smooth boundaries instead of rough boundaries, where smooth boundaries absorb $m$ particles instead of $e$ particles. Indeed, protocols to implement quantum logic gates have been shown which involve holes with both types of boundary~\cite{Raussendorf06, Raussendorf07, Raussendorf07a}. It is also worth noting the recent work of Delfosse~{\it et al.}~\cite{Delfosse16} where they study the encoding properties of holes which have both rough and smooth boundaries. Here we need only introduce holes with rough boundaries. In what follows we will see that rough boundaries differ from smooth boundaries only by lattice defects that are introduced in the following Subsection.

\subsection{Encoding logical qubits with twist defects}
\label{subsctn:PlanarCodeWithTwists}

We next consider encoding qubits using twist defects. Twist defects were introduced for the surface code model by Bomb\'{i}n in Ref.~\cite{Bombin10}. Other work on twist defects on the surface code is presented in Refs.~\cite{Kong12, Brown13, Wootton15a, Zheng16, Teo16}. Twist defects have also been introduced in the color code models in Ref.~\cite{Bombin11, Teo14}, and from the point of view of topological quantum field theories in Refs.~\cite{BarkeshliJianQi13, Barkeshli14} and references therein. See also the second item in Section~12 of Ref.~\cite{Kitaev06}.

\begin{figure}
\includegraphics{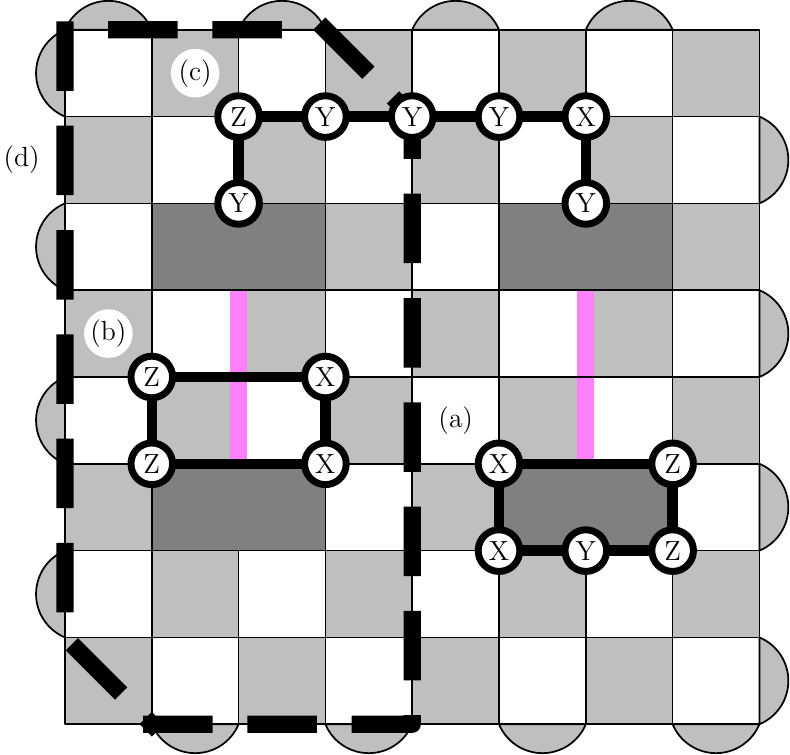}
\caption{Two pairs of twist defects on the planar code. Twists appear at the end points of the pink defect lines. (a)~The weight-five stabilizer operator of a twist defect, shown up to a phase factor. (b)~A modified weight-four stabilizer that lies on the pink line. (c)~A logical operator associated to the four twist defects on the lattice. (d)~The tensor product of Pauli-X operators on the qubits supported along the dashed line gives the logical operator that anti commutes with that shown at (c).\label{fig:FourTwists} }
\end{figure}

In Fig.~\ref{fig:FourTwists} we show four twists on a lattice which collectively encode a single qubit. On this lattice the qubits that lie along the two pink lines are removed from the lattice. We refer to these pink lines as defect lines. Twist defects lie on the double plaquettes where the defect lines terminate. Up to a complex phase, we show a stabilizer that lies on a twist defect in Fig.~\ref{fig:FourTwists}(a). Indeed, one should also include a complex $\pm \text{i}$ phase in front of the twist stabilizer such that the stabilizer returns a $+1$ measurement outcome, where the sign determines the phase the defect carries. We will largely neglect this phase, as is does not significantly change the physics that we are demonstrating, but we direct the interested reader to Ref.~\cite{Bombin10}. In Fig.~\ref{fig:FourTwists}(b) we also show how the weight-four stabilizers are modified along the defect line. In general, the stabilizers that lie on the defect lines are determined as follows; we first replace the stabilizers that lie along defect lines with restricted stabilizers where we take the restriction~\footnote{Precisely, given an operator $O = O_A \otimes O_B $, the restriction of $O$ onto $A$ is $O_A \otimes \openone_B$.} on the qubits that do not support defect lines. We then replace the restricted stabilizers with the products of pairs of adjacent restricted stabilizers. Indeed, following this prescription gives the stabilizers shown in Figs.~\ref{fig:FourTwists}(a) and~(b). We show the logical operator $\overline{X}$ in Fig.~\ref{fig:FourTwists}(c), and the logical operator $\overline{Z}$ is the tensor product of Pauli-X operators on all the qubits that support the dashed line in Fig.~\ref{fig:FourTwists}(d). If we have a single uniform lattice boundary, we can see from the logical operators that to maintain a distance $d$ code, all twist defects must be separated by $\sim d$, see Ref.~\cite{Hastings15}.

We now briefly review the properties of twists from the point of view of anyonic quasiparticles. Indeed, the string operators we have studied create either a pair of electric charges or magnetic charges at the two end-points of the string when they act on codewords in the bulk of the lattice. In this sense we see that the surface code model obeys global charge parity conservation law as anyonic charges of the same type must be created in pairs. Interestingly, string operators that cross a defect line create one $e$ excitation and one $m$ excitation at its end points. Remarkably, it follows from this fact that this property gives twist defects the ability to absorb a fermion. It was observed by Bomb\'{i}n in Ref.~\cite{Bombin10} that this behaviour is reminiscent of Ising anyons, which we reviewed in Subsubsec.~\ref{Subsubsctn:Ising}. Importantly, like twists, Ising anyons also have the ability to absorb fermionic excitations.

We can see that twist defects mimic the behaviour of Ising anyons by consideration of their logical operators. Indeed, the string of Pauli-Y operators shown in Fig.~\ref{fig:FourTwists}(c) represents a string operator that transports a fermion from one twist on the lattice to another. The operator $\overline{Z}$ on the other hand measures the fermionic charge parity that has been absorbed by the two twist defects that lie on the left of the lattice. With this observation we see that the physics of Ising anyons is echoed by the simple stabilizer model that we consider here. Later we make use of this analogy to find new logical gates by code deformation, which we discuss in the following Section.

\subsection{A twist on the boundary}
\label{subsctn:TwistsOnBoundary}
Before moving onto the next Section, we finally demonstrate a correspondence between twist defects and the corners of the planar code. More precisely, by corners we mean points on the boundary of the planar code where rough boundaries meet smooth boundaries. In later Sections we make use of this correspondence to perform Clifford gates with the planar code using code deformations.

\begin{figure}
\includegraphics{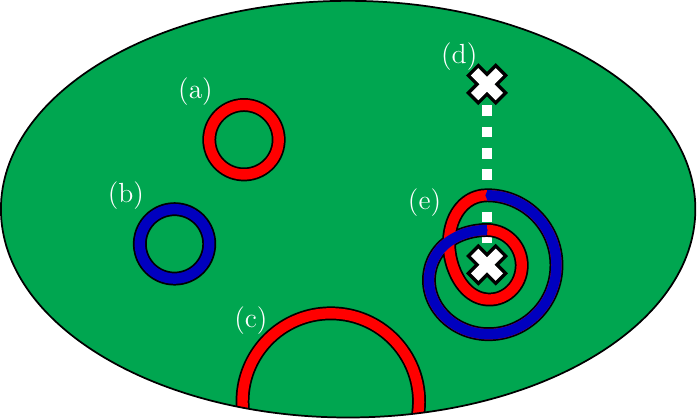}
\caption{\label{Fig:AnyonDiagram} A schematic diagram of the planar code lattice with a rough boundary. Strings of Pauli-Z(Pauli-X) operators are marked by red(blue) lines, respectively. (a)~and (b)~Respectively show $A_f$ and $B_f$ stabilizer operators. (c) An element of the stabilizer group where a string of Pauli-Z operators follows a trivial path and terminates at the same rough boundary of the lattice. (d) A defect line marked by a dashed white line with two twist defects at its end points. Red strings that cross the defect line change to blue and vice versa. (e) A stabilizer operator that encloses a single twist. The string must wind twice around the twist which changes the string from red to blue and back to red again to form a closed loop. As mentioned, the stabilizer should also include a complex phase due to the point where the two strings cross, which we do not consider explicitly.}
\end{figure}

To show the correspondence between twists and corners, we will consider the planar code lattice with defect lines in different configurations. To do so, we first build a diagrammatic notation without the microscopic details of the underlying lattice.

 In Fig.~\ref{Fig:AnyonDiagram} we show a planar code with a rough boundary that supports a single pair of twist defects. In the diagram, strings of Pauli-Z(Pauli-X) matrices are shown by red(blue) strings. With this picture, the operator $A_f$($B_f$) can be regarded as a red(blue) string that follows a trivial cycle, as shown in Figs.~\ref{Fig:AnyonDiagram}(a) and~(b), respectively. As the lattice has only a single rough boundary, the stabilizer group also contains strings of Pauli-Z operators that follow trivial cycles and terminate at the boundary of the lattice, as shown in Fig.~\ref{Fig:AnyonDiagram}(c).  Importantly, as the boundary is entirely rough, only red strings can terminate at the boundary.  We show a defect line in Fig.~\ref{Fig:AnyonDiagram}(d), marked by a thick dashed white line, where two twist defects lie at its termination points. Twists are shown as white crosses. We also show an example of a stabilizer operator that encloses a single twist in Fig.~\ref{Fig:AnyonDiagram}(e) drawn as strings of Pauli-X and Pauli-Z operators, where the string changes colour as it crosses the boundary.

\begin{figure}
\includegraphics{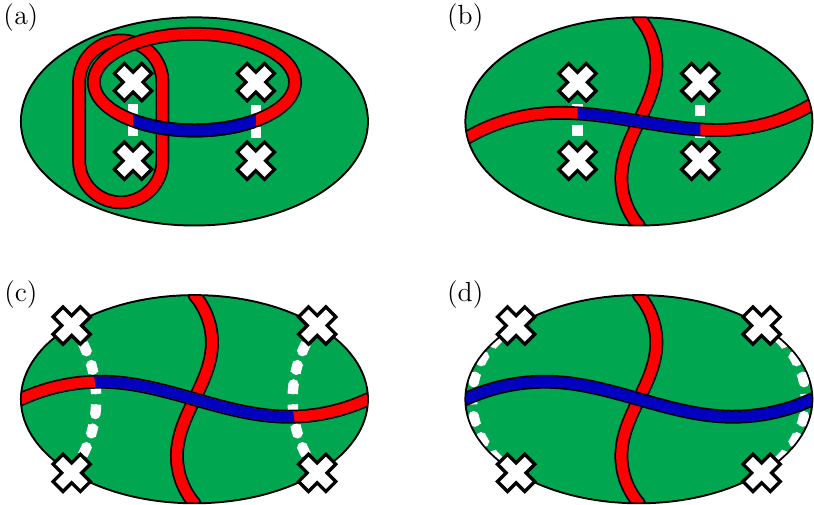}
\caption{Lattices with different configurations of twist defects but equivalent logical operators. (a)~Four twist-defects in the centre of the lattice. The logical operators are closed loops that enclose pairs of twists. (b)~The logical operators shown in~(a) mulitplied by stabilizer operators which terminate at the boundary, as in Fig.~\ref{Fig:AnyonDiagram}(c), thus giving the logical operator shown in the picture. (c)~We consider the lattice shown in~(a) and~(b), except where the defect lines terminate at the boundary. In this picture the twist defects lie at the boundaries of the lattice. Importantly, the logical operators are unchanged from those shown in~(b). (d)~We finally show the lattice where the defect lines are drawn close to the left and right sides of the lattice. Now, the logical operators are equivalent to those in Fig.~\ref{fig:WPM} where a string of Pauli-X operators runs from the left to the right of the lattice and a string of Pauli-Z operators runs from the top to the bottom of the lattice. 
  \label{Fig:CornersAreTwists}}
\end{figure}

Having introduced a simple diagrammatic notation, we can easily demonstrate the equivalence between the corners of the planar code and twists which we summarise in Fig.~\ref{Fig:CornersAreTwists}. In Fig.~\ref{Fig:CornersAreTwists}(a) we show a lattice that encodes one logical qubit using four twist defects. The logical operator $\overline{X}$ is a string operator that encloses the top two twists. As previously discussed, this operator corresponds to an operator that transports a fermion between the two enclosed twists. We see this because the parallel strings are passing an $e$ and an $m$ excitation between the two twists. The operator $\overline{Z}$ is shown by a red loop that encloses the two twists to the left of the figure. This measures the parity of fermonic charges that have been absorbed by the two enclosed twists. From the diagram we can see that the two logical operators anti commute because there is one single point where the blue segment of $\overline{X}$ intersects the red $\overline{Z}$ loop.

Logical operators can be manipulated by multiplying them by stabilizer operators. As we have discussed stabilizer operators are strings of trivial cycles, or strings that both terminate at the same boundary. In Fig.~\ref{Fig:CornersAreTwists}(b) we show the logical operators of Fig.~\ref{Fig:CornersAreTwists}(a) where the logical operators are deformed such that they terminate at the boundary of the lattice. As with Fig.~\ref{Fig:CornersAreTwists}(a) we see these two logical operators anti commute as there is still a single point where the blue segment of the $\overline{X}$ operator intersects the red string that represents the $\overline{Z}$ operator.

Indeed, the logical operators in Fig.~\ref{Fig:CornersAreTwists}(b) are reminiscent of the logical operators of the planar code discussed in Subsec.~\ref{subsec:surfcode}, where logical operators also terminate at the boundary. We next consider the same model, except where the defect lines terminate at the boundary of the lattice, such that the four twist defects of the code are located at the boundaries of the lattice. This is shown in Fig.~\ref{Fig:CornersAreTwists}(c). In this picture, the logical operators are unchanged from those shown in Fig.~\ref{Fig:CornersAreTwists}(b), which, once again, are logical operators that are very similar to those of a planar code. 
To make this analogy completely clear, we show the same model in Fig.~\ref{Fig:CornersAreTwists}(d), except where the defect lines have been drawn close to the boundary of the lattice, but where the twist defects remain in the same locations. In this picture, the logical operators are now identical to those of the planar code, where $\overline{Z}$ corresponds to moving an $e$ excitation from the upper boundary to the lower boundary, and $\overline{X}$ corresponds to moving an $m$ excitation from the left boundary to the right boundary. In this sense, we can regard a smooth boundary as a rough boundary that has been covered by a defect line. We observe also that the points, or corners, where the rough boundary meets the smooth boundary are the locations of twist defects. We can therefore regard the corners of the planar code as equivalent to twist defects. In the following Section, we will make use of this analogy to demonstrate how we can perform logical Clifford gates on the planar code by code deformations that manipulate the corners of the planar code lattice as though they are twist defects.

\section{Logical operations by manipulating corners}
\label{Sctn:SingleQubitGates}

We have now introduced several different methods of encoding qubits in the surface code model. However, quantum computation requires that we also perform logical gates on encoded qubits. A well-studied method for performing fault-tolerant quantum logical operations is by use of code deformations~\cite{Raussendorf06, Raussendorf07, Raussendorf07a, Bombin08, Bombin09, Fowler09, Fowler12a, Brell15, Cong16, Cong17}, where we make special measurements to manipulate the logical qubits of a quantum error-correcting code. In particular, we are going to perform code deformations to manipulate and braid the twist defects that lie at the corners of the planar code, as discussed in the previous Section. Given that the twist defects of the planar code are analogous to Ising anyons, we can devise code deformation strategies based on known braiding gates that can be achieved using Ising anyons~\cite{Bravyi06, Kitaev06}. The manipulation of twist defects has been discussed in the following references~\cite{Bombin10, Bombin11, You12, Teo14, Wootton15a}. Indeed, like Ising anyons, twist defects can achieve the Clifford group by braiding which, together with magic state distillation~\cite{BravyiKitaev05}, can be used to achieve universal quantum computation. In this Section we begin by reviewing the theory behind code deformations. We then show explicitly how to move twists by code deformations. We finally show how to perform single-qubit Clifford gates at the end of this Section.

\subsection{Code deformations}

Here we briefly review the concept of code deformation. This is a process by which a code defined with stabilizer group $\mathcal{S}$ is mapped onto a different code defined by stabilizer group $\mathcal{S}'$. With a suitable choice of code deformation we can perform logical rotations on a code. Code deformation is achieved simply by measuring the elements of $\mathcal{S}'$ to project onto the new code. Details on performing measurements using the stabilizer formalism are given in Chapter~10 of Ref.~\cite{Nielsen00}. We add that the code deformation procedures we present should be readily adapted to adiabatic topological quantum computational schemes using the methods shown in Refs.~\cite{Bacon10, Cesare15}.

We require that no stabilizers of $\mathcal{S}'$ measure the logical information encoded in the code space of the code specified by $\mathcal{S}$ to preserve encoded information. For all of the instances we consider here, we can perform sequential measurements with weight that is much smaller than the distance of the code. It then follows that we can always clean logical operators~\cite{Bravyi09, Bravyi11} away from the qubits that support the stabilizer measurements of $\mathcal{S}' $. We therefore focus on how the stabilizer group is modified by the stabilizer measurements of $\mathcal{S}'$ on the code specified by $\mathcal{S}$. 

We consider the simple case where the generators of $\mathcal{S}'$ differ from $\mathcal{S}$ by only a single element, $s'$. This is easily generalised to the case where $\mathcal{S}'$ differs from $\mathcal{S}$ by many elements, as we can sequentially deform the code several times between many different codes. Depending on the choice of $\mathcal{S}'$, element $s'$ will do one of two things; it will either commute with all elements of $\mathcal{S}$, or it will anti commute some stabilizers of  $\mathcal{S}$. In the case that $s'$ commutes with all elements of $\mathcal{S}$, provided $s'$ is not a logical operator, then it must follow that $s' \in \mathcal{S}$, which is a trivial deformation. 

We next consider the case where $s'$ anti commutes with some elements of $\mathcal{S}$. We denote the subset of elements of $\mathcal{S}$ that do not commute with $s'$ as $\mathcal{A}$, where we denote members of the anti commuting set $s_j \in \mathcal{A}$ where $ 1 \le j \le N$ and $N$ is the number of elements in $\mathcal{A}$. In this case, we replace elements of $\mathcal{A} \subseteq \mathcal{S} $ with terms of the form $s_j s_{j+1}$ for all $ 1 \le j \le N-1$. Then, up to the measurement outcome $m' = \pm 1$, we project our encoded state onto the codespace of $\mathcal{S}'$. In the case that $ m'  = - 1$, we project the encoded state onto the $-1$ eigenstate of $s'$. If this is so then we can apply a unitary correction operator to rotate the state we achieved under the projection onto the desired state.

\subsection{Manipulating corners by code deformations}

\begin{figure}
\includegraphics{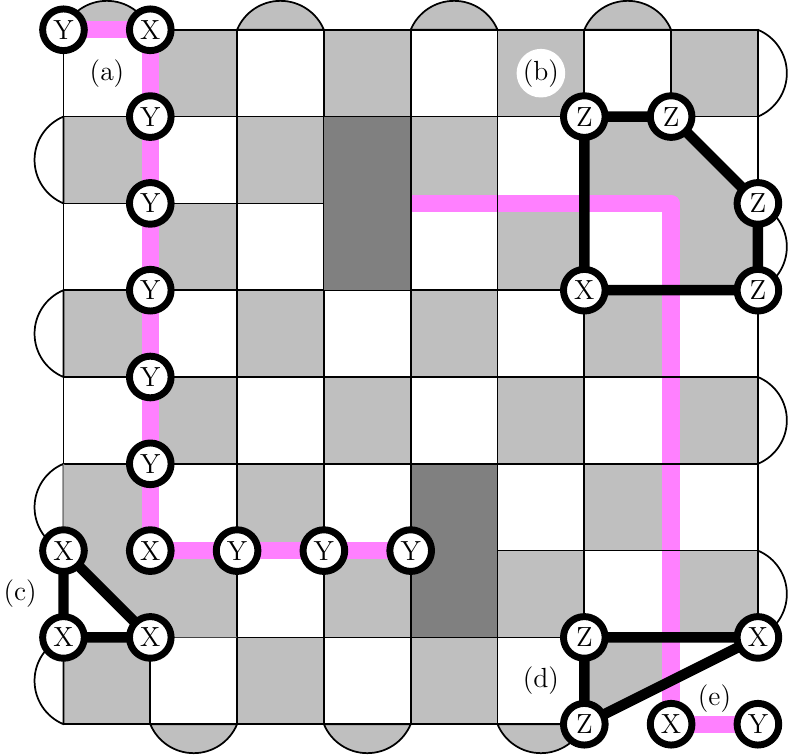}
\caption{Moving the twist defects at the corners of the planar code into the bulk of the lattice. (a)~We make single-qubit Pauli measurements along the pink defect line to deform the twist defect from the corner of the planar code to the centre of the lattice. Most of the measurements along the line are Pauli-Y operators, but we also use Pauli-X measurements and Pauli-Z measurements to move a twists around a corner. (b)~and~(c) Stabilizer operators that are modified by the code deformation at locations where twists are moved around corners. (d)~The stabilizer operator that lies at the corner of the lattice where a twist defect has been removed. (e)~Two qubits at the corner of the lattice that have been measured in a product basis and are no longer entangled to the code. \label{fig:DeformingCorners}}
\end{figure}

We now show that we can move the corners of the planar code onto the bulk of the lattice. The sequence of measurements we choose are based on the terms in the perturbation expansion used to synthetically introduce twist defects to Hamiltonian models~\cite{You13}.

In Fig.~\ref{fig:DeformingCorners}(a) we show the single-qubit Pauli measurements that we perform along the pink line to move the twist into the bulk of the lattice. Upon performing these measurements, the twist defect is moved along the pink line. After performing these measurements, the qubits that lie on the pink line are projected onto a product state, and are thus disentangled from the lattice.

With a few exceptions which we show in Fig.~\ref{fig:DeformingCorners}, the stabilizers along the pink defect line of the deformed lattice are found by the same prescription given in Subsec.~\ref{subsctn:PlanarCodeWithTwists}. Twist defects are weight-five stabilizers at the terminal point of the defect line, and stabilizers along the straight segments of the defect line are weight-four operators that straddle the line. We also consider how the stabilizers are modified in the locations where the defect line moves around a corner. In Figs.~\ref{fig:DeformingCorners}(b) and~(c) we show some explicit examples of one weight-three stabilizer, and one weight-five stabilizer where the twist defect turns around a corner on the lattice. We also show how the stabilizers are modified close to the corner where the twist defect originated in Figs.~\ref{fig:DeformingCorners}(d) and~(e). In Fig.~\ref{fig:DeformingCorners}(d) we show a weight-three stabilizer on the boundary of the lattice where the twist defect began, and in Fig.~\ref{fig:DeformingCorners}(e) we show two qubits that have been projected into the product state, and are thus removed from the code.

\subsection{Single-qubit Clifford rotations on the planar code by code deformation}
\label{Subsctn:SingleQubitGates}

\begin{figure}
\includegraphics{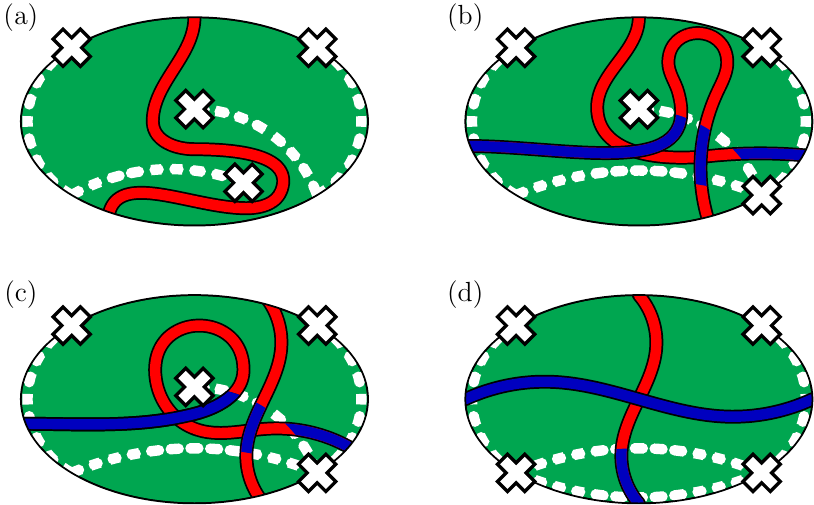}
\caption{A sequence of code deformations that map the logical operator $\overline{Z}$ onto $\overline{Y}$. (a)~We move the bottom two twist defects into the centre of the lattice. We also show the $\overline{Z}$ string operator in red which `snakes' between the twist defects at the centre of the lattice. (b)~We deform the position of one twist defect into the bottom-right corner of the lattice. The logical operator is deformed over the defect line such that it is not supported on any of the qubits where code deformation measurements which move the lowest twist defect from its position in~(a) to its position in~(b). This deformation moves the logical operator over a defect line, and as such changes the colour of the string. We also multiply the logical operator by a stabilizer that extends from the left-hand side of the lattice to the bottom of the lattice, which is also shown in the figure. (c)~A small deformation of the logical operator shown in~(b) gives the pictured logical operator. (d)~Multiplying the logical operator shown in~(c) by a stabilizer that encloses the single twist defect lying at the centre of the lattice deforms the logical operator out of the path of the twist, enabling us to move the twist defect at the centre of the lattice to the bottom left corner of the lattice, thus completing the procedure. 
\label{fig:PerformingS}}
\end{figure}

Having demonstrated that it is possible to move the corners of planar codes into the bulk of the lattice and braid them like twist defects, we now show that we can complete the generating set of the Clifford group using twist defects. In terms of scope and resource usage, our scheme can be compared with protocols given in Ref.~\cite{Fowler12a, Fowler12b} where Hadamard gates are performed on qubits encoded with pairs of holes.

In Fig.~\ref{fig:PerformingS} we show that a deformation that exchanges two twists on two adjacent corners maps between different Pauli matrices. Specifically, we show that exchanging the two twists at the bottom of the lattice will, up to phases, exchange logical operators $\overline{Z} \rightarrow \overline{Y}$ and $\overline{Y} \rightarrow \overline{Z}$, and where $\overline{X}$ remains invariant. We work through this manipulation step by step by manipulating the support of the logical operators shown in Subsec.~\ref{subsctn:TwistsOnBoundary}. The strategy we will follow will be to deform the logical operator away from the path of the twist defects. 

Note that, up to phases, the effect of this exchange is equivalent to the braiding of two Ising anyons~\cite{Bravyi02, Bravyi06}. Indeed, as noted in Subsection \ref{subsec:anyons}, exchanging two Ising anyons is equivalent to the square root of the $\overline{X}$ operation associated with these twists.

\begin{figure}[b]
\includegraphics[width=\columnwidth]{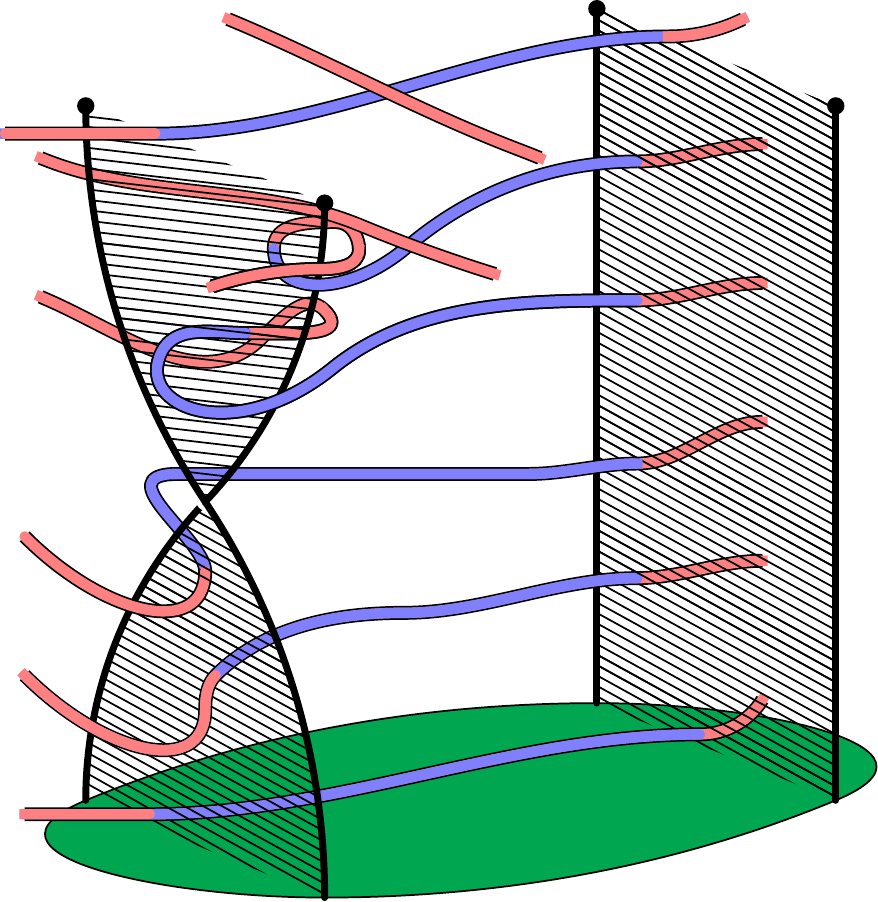}
\caption{Spacetime figure showing the evolution of the logical operator $\overline{X}$ under the exchange of two twists, where the time direction is upwards along the page. The figure shows the operator is mapped onto the logical operator $\overline{Y}$, which can be seen at the top of the figure. The trajectory of the twist defects are marked by black lines. As in the two-dimensional figures above, red(blue) strings correspond to the world lines of $e$($m$) quasiparticle excitations. We continue with the convention we used earlier, where $e$ charges can terminate at the boundary. We depict this by showing red strings diverging away from the twist defects. \label{Fig:Spacetime}}
\end{figure}

We first deform the two twists on the corners of the lower boundary into the centre of the lattice, as shown in Fig.~\ref{fig:PerformingS}(a). Upon doing this we deform the logical operator $\overline{Z}$ away from the positions of the twists by multiplying the logical operator by elements of the stabilizer group such that none of the qubits that are measured during the code deformation support the logical operator. The logical $\overline{Z}$ operator is shown in red in Fig.~\ref{fig:PerformingS}(a) snaking between the twists in the centre of the lattice.

Next, we continue to move the twist defect that began in the bottom-left corner of the lattice into the bottom-right corner of the lattice. We therefore continue to deform the logical operator such that one end point terminates on the right-hand side of the lattice. This forces the logical operator to move over a defect line which has already been drawn on the lattice, which changes its string type from red to blue, as we show in Fig.~\ref{fig:PerformingS}(b). We also multiply the logical operator by an element of the stabilizer group which extends from the left boundary to the lower boundary of the lattice. Introducing this stabilizer operator at this point will help us see that we can deform the logical operator out of the trajectory of the twist defect that lies in the centre of the code as it moves towards the bottom-corner of the lattice.

In Fig.~\ref{fig:PerformingS}(c), we show a small deformation of the logical operator shown in Fig.~\ref{fig:PerformingS}(b), where now we have the product of a string operator running from the top to the bottom of the lattice, and a string that runs horizontally across the lattice that loops around the one twist defect that remains in the bulk of the lattice. To deform the logical string shown in Fig.~\ref{fig:PerformingS}(c) we multiply the logical operator by a stabilizer that loops around a single twist, as in Fig.~\ref{Fig:AnyonDiagram}(e), to deform the horizontal string over the central twist. This allows us to move the central twist defect into the bottom-left corner of the lattice without performing any code deformation measurements over the support of the logical operator. Then, recognising that the two parallel defect lines that run along the bottom edge of the lattice are equivalent to a boundary where there is no defect line, we recover the lattice shown in Fig.~\ref{Fig:CornersAreTwists}(d).

We finally look at the action of the deformation on the $\overline{X}$ operator. One can readily check that this logical operator is invariant under the presented transformation, as this logical operator can be supported on the qubits on the lattice that are never acted upon by non-trivial measurements under the code deformation procedure, thus showing the promised action of this code deformation procedure.

To complete the Clifford group, one can also check using a similar argument that, up to phases, exchanging the two twists at the left hand side of the lattice will map the Pauli operators such that $\overline{X} \rightarrow \overline{Y}$, $\overline{Y} \rightarrow \overline{X}$, and will leave the operator $\overline{Z} $ unchanged. For an alternative perspective, we show this operation in a spacetime diagram in Figure~\ref{Fig:Spacetime}. Now, if we denote the exchange of the two twists at the bottom(left-hand side) of the lattice as $B_1$ ($B_2$), then, up to Pauli-rotations, we have the single-qubit logical phase gate $\overline{S}$ and the logical Hadamard gate, $\overline{H}$ such that $B_2 = - \overline{X}\cdot \overline{S}$, and $B_1 B_2 B_1 = - \overline{Y} \cdot \overline{H}$ which generate the single-qubit gates of the Clifford group. Given that we can achieve logical Pauli-matrices either via transversal single-qubit Pauli rotations, or, more simply, by updating the Pauli-frame~\cite{Fowler12a}, we recover a fault-tolerant implementation of the Clifford group with code deformation using the planar code.

\begin{figure}
\includegraphics{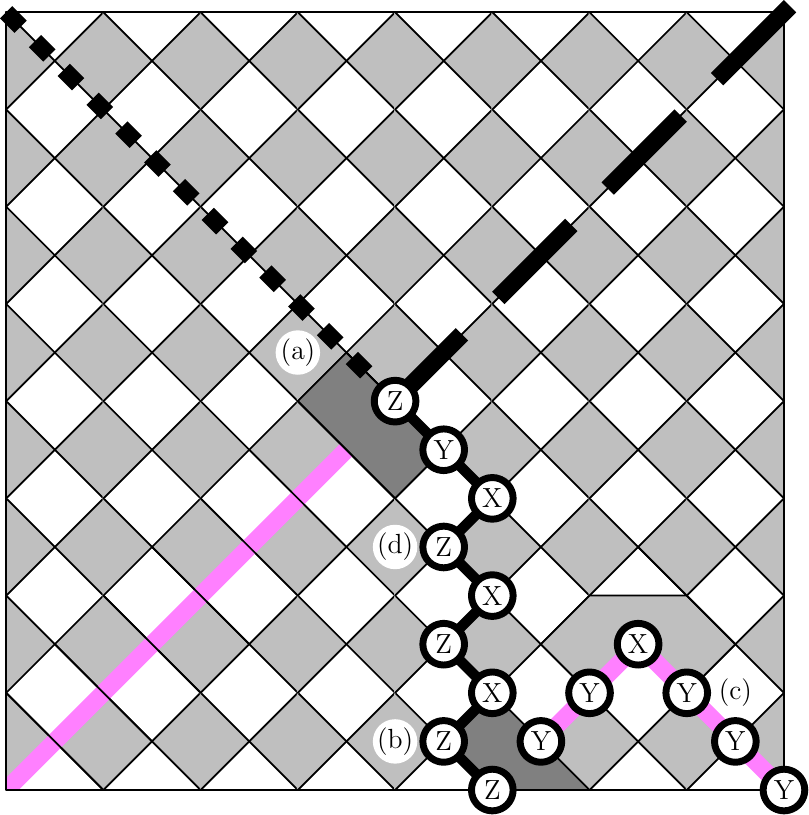}
\caption{By rotating the lattice geometry about an angle of $\pi / 4$ we can perform both twist exchange operations without reducing the distance beyond a small constant value. Figure shows a distance $L = 9$ lattice. (a)~and (b)~Two twist defects that have been deformed into the centre of the lattice. (c)~The single-qubit measurements we make to move the twist along the bottom of the lattice. (d)~A weight-nine logical operator at a moment where the two twists pass each other. The black dashed and dotted lines show the support of conjugate weight-nine logical operators that we find to be minimal in weight.
\label{fig:DiamondLattice}}
\end{figure}

Upon performing the suggested braiding operations with the square lattice introduced in the previous Section we modify the code distance. Recalling that the distance of a logical qubit encoded via twist defects depends on the separation between all the twists on the lattice, it is obvious that we must deform the locations of the twist defects such that they all maintain a large separation. Using the lattice shown in Fig.~\ref{fig:WPM}, we can find paths that the twists follow while undergoing this exchange such that all the twist defects maintain a distance of at least $\sim L/2$ from one another on the considered lattice geometry. As such, the code distance is $d \sim \mathcal{O}(L/2)$ as we perform the presented logical operations. Interestingly, we also find that the code with a rotated geometry, as is shown in Fig.~\ref{fig:DiamondLattice}, that we can exchange the corners without more than a small constant loss in code distance. This comes at the expense of using $2d^2$ qubits to achieve a code of distance $d$, which is in contrast to the square geometry we have already introduced which requires only $d^2$ qubits. It is interesting then that the resource demands of both lattice geometries are similar when we consider performing logical gates by braiding the corners of the lattice.

\subsection{Encoding two qubits on a single planar code} 
\label{subsec:doublediamond}

By regarding the corners of the planar code as twist defects, we have seen that we can braid them to realize a set of gates that generates the single-qubit elements of the Clifford group. We can extend this idea by adding more corners to the planar code, such that it encodes more qubits. Then, we can braid the corners of this extended code to implement entangling gates, as we describe in this Subsection. We believe that the entangling operation we present here may provide a relatively simple procedure to entangle two fault-tolerant qubits in the laboratory using near-future technology.

\begin{figure}
\includegraphics{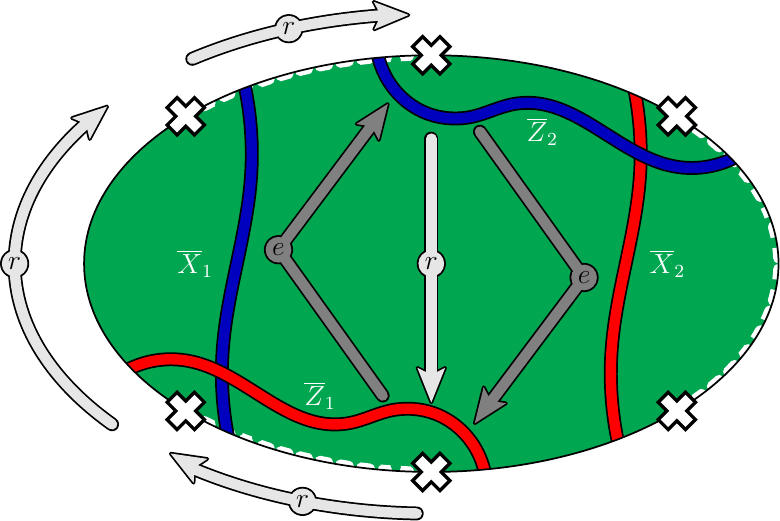}
\caption{A surface code which encodes two logical qubits. Logical operators $\overline{X}_1,\, \overline{Z}_1,\, \overline{X}_2$ and $\overline{Z}_2$ which act on qubit $1$ and qubit $2$ are shown. Exchanging the two central qubits along the trajectories shown by the dark grey arrows which are marked by $e$ entangle the two logical qubits. This exchange can be performed without decreasing the code distance beyond a small constant amount on an $L \times 2L$ lattice with the square geometry we show in Fig.~\ref{fig:WPM}. \label{fig:DoubleDiamond} }
\end{figure}

We look at the example shown in Fig.~\ref{fig:DoubleDiamond} which encodes two logical qubits which we index qubit $1$ and qubit $2$. In the picture we show a planar code with six twist defects lying on the boundary of the surface code. The two twist defects at the left of the figure and the twist defect at the bottom centre of the lattice support the logical qubit $1$, and the other three twist defects support qubit $2$. The logical operators, $\overline{X}_1,\, \overline{Z}_1,\, \overline{X}_2$ and $\overline{Z}_2$, are shown in the Figure. As in the previous Subsection, we can generate the single-qubit Clifford operations by exchanging the twist defects of a respective qubit. 

Additionally, we can also perform an entangling operation between the two qubits by exchanging the two twists in the centre of the lattice.

Each pair of twists is individually able to fuse to either vacuum of a $\psi$ particle. However, for any state within the stabilizer space, the parity of the number of $\psi$ particles overall must be even. This is to ensure that they can mutually fuse to vacuum. Given this restriction, the fusion result for the leftmost and rightmost pairs must always be the same as that of the central pair. This pair is therefore associated with the logical operator $\overline{X}_1 \overline{X}_2$. Consider exchanging these two twists along the dark grey arrows shown in Fig.~\ref{fig:DoubleDiamond} that are marked with a letter $e$. Explicitly, we find that under the exchange of the central two twists gives rise to the transformation
$$
\overline{X}_1 \rightarrow \overline{X}_1, \quad \overline{Z}_1\rightarrow \overline{Y}_1 \cdot \overline{X}_2,
$$
$$
\overline{X}_2 \rightarrow \overline{X}_2, \quad \overline{Z}_2 \rightarrow \overline{X}_1 \cdot \overline{Y}_2,
$$
which can be shown using methods similar to those used in Subsec.~\ref{subsec:anyons}. This gate is equivalent to a controlled-not gate up to local Clifford rotations.

Remarkably we can perform both the single-qubit operations, and the entangling gate, without decreasing the distance of the code by using an $ L \times 2L $ lattice with square geometry as shown in Fig.~\ref{fig:DiamondLattice}, where we have $\sim 2d^2$ physical qubits describing two logical qubits with a distance $\sim d$ code. As such, we believe, given the recent surge in progress in experimental quantum error correction~\cite{chiaverini04, reed12, nigg14, Corcoles15, Kelly15, Takita16, Ofek16, Albrecht16}, that this example provides a relatively simple experiment to demonstrate the full Clifford group that may be implemented in the near future.

Note also that by exchanging the leftmost and middle twists using the light grey arrows marked $r$, and doing similar for the right hand side, we can achieve the direct exchange of any twist pair in the center without loss of distance. The equivalent gates can be achieved without using such rotations, but it could yield superior results in certain cases by keeping the twists slightly better separated.

\section{Entangling different types of logical qubits}
\label{Sctn:EntanglingGates}

In addition to single-qubit elements of the Clifford group, it is also important to have fault-tolerant schemes that entangle many logical qubits. It is known that we can perform entangling operations using two-qubit non-destructive parity measurements~\cite{Terhal15}. One method of performing a two-qubit parity measurement non-destructively is to prepare a third ancillary qubit which is then entangled to the two qubits we wish to perform a parity measurement over, and then measure the ancilla qubit. In this Section we elaborate upon one of the fault-tolerant entangling schemes in Ref.~\cite{Hastings15} which performs a controlled-not gate between two qubits that are encoded using quadruples of twists using a third ancillary logical qubit. We first show in detail that we can perform a controlled-not gate between a logical qubit encoded with a pair of holes and a logical qubit encoded over four twist defects by code deformations. We can then use this entangling gate to perform parity measurements between qubits encoded with twists by performing entangling gates between logical qubits and an ancillary qubit encoded with a pair of holes, and subsequently measuring the ancilla qubit. We use this entangling operation to perform a controlled-not gate between qubits encoded with twist defects. We point out that this scheme is much of a likeness to a measurement-only topological quantum computation scheme~\cite{Bravyi06, Zilberberg08, Bonderson08, Bonderson09, Levaillant15, Freedman15, Zheng16} presented by Bravyi in Ref.~\cite{Bravyi06}.

\subsection{Braiding twists and holes}
\label{Subsctn:EnganglingTwistsAndHoles}
\begin{figure}
\includegraphics{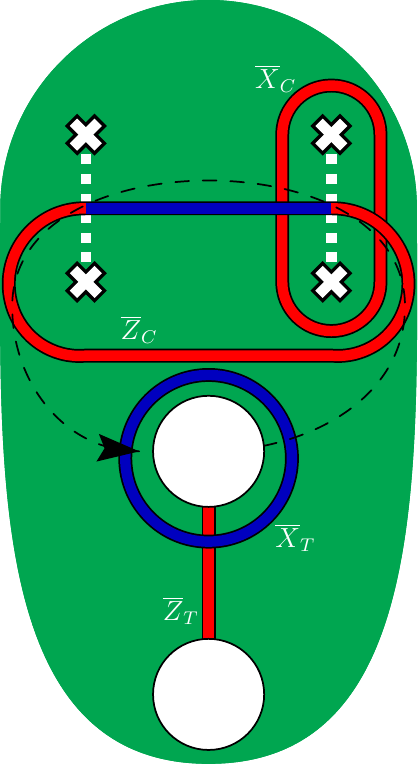}
\caption{\label{fig:OneTwistAndOneHoleQubit} The code deformation scheme to entangle a qubit encoded with a pair of holes to a qubit encoded over four twists. At the top of the figure we show the logical operators for a quadruple twist qubit encoding, and at the bottom of the figure we show the logical operators for a hole-pair qubit encoding. We present a scheme to perform a controlled-not gate where the qubit encoded with the four twists is the control qubit and the qubit encoded using holes is the target qubit. The black dashed line shows the trajectory the hole in the middle of the figure must follow to complete a controlled-not gate between these qubits.}
\end{figure}

We consider the setup shown in Fig.~\ref{fig:OneTwistAndOneHoleQubit}. In the picture we show two qubits, one encoded over four twists on the lattice, and the other encoded using two punctures.
We call a qubit encoded with four twists the quadruple-twist encoding, and a qubit encoded using two punctures the hole-pair encoding.
We show that deforming one hole along a path that encloses two twists, whose trajectory we show with a dashed line in the figure, will execute a controlled-not gate. We show the action of the braid operation by demonstrating that tthe logical operators shown in Fig.~\ref{fig:OneTwistAndOneHoleQubit} satisfy the conditions given in the Eqns.~(\ref{eqn:CNOT}).

We will not elaborate in detail the sequence of code deformation measurements we perform to move holes, as this has been described in detail in, for instance, Refs.~\cite{Raussendorf06, Raussendorf07, Raussendorf07a, Bombin08, Bombin09, Fowler09, Fowler12a}. For now, it is enough to understand that holes are moved by first increasing their size along some suitable direction, and then decreasing their size again, such that the hole is displaced along their path of motion. More specifically, we increase the size of a hole with a rough(smooth) boundary by measuring the physical qubits of the lattice with single-qubit Paui-Z(Pauli-X) close to the boundary of the puncture.  As such, the qubits in the centre of a hole with a rough(smooth) boundary are measured onto an eigenstate of the Pauli-Z(Pauli-X) matrix, see Fig.~\ref{fig:Hole}(c). To decrease the size of the hole, we measure stabilizers at the boundary of the puncture. Importantly, as we perform measurements to move the hole, we will not change its topology. We can continually repeat this prescription to transport a hole along the different trajectories described below.

\begin{figure}
\includegraphics{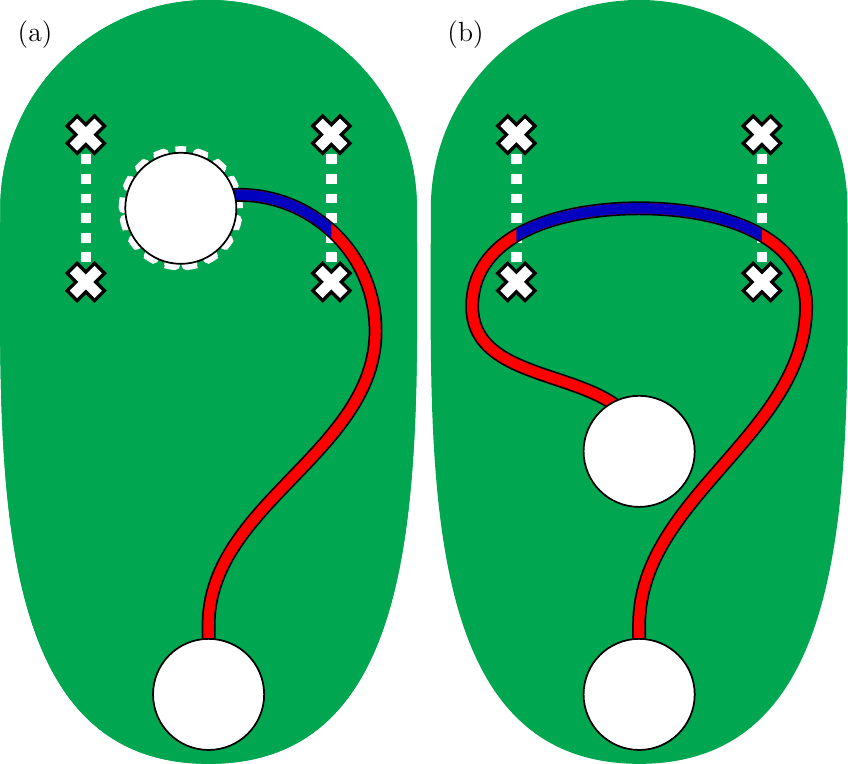}
\caption{\label{fig:ZtoZZ} The deformation of the $\overline{Z}_T$ operator under the fault-tolerant entangling operation. (a)  The logical operator that terminates at the hole moves with the hole, and is thus threaded between the twists. We also point out that the boundary of the hole changes from a rough boundary to a smooth boundary as it is passed across the defect line. As such, we draw a white dashed line around the puncture. (b) Upon moving the puncture back to its initial position we see that we have wrapped the logical string operator around the lower two twists shown in the figure. One can check that the deformed logical operator is equivalent to $\overline{Z}_C\overline{Z}_T$, as they are shown in Fig.~\ref{fig:OneTwistAndOneHoleQubit}. }
\end{figure}

Having discussed how holes are moved around the lattice, we next show that performing the braid shown in Fig.~\ref{fig:OneTwistAndOneHoleQubit} will perform a controlled-not gate. We first show that $\overline{Z}_T \rightarrow \overline{Z}_C \overline{Z}_T$. This is shown diagrammatically in Fig.~\ref{fig:ZtoZZ}. We consider the transformation of the $\overline{Z}_T$ that is shown in Fig.~\ref{fig:OneTwistAndOneHoleQubit}. As the hole moves, the logical operator stretches to follow the hole. This causes the logical operator to thread between the twists, as we show in Fig.~\ref{fig:ZtoZZ}(a). We note that as the hole passes across a defect line it changes its boundary type such that it now absorbs Pauli-X string operators instead of Pauli-Z string operators. Once the deformation procedure is completed and the hole is returned to its initial position, the logical operator is deformed around the lower two twists on the lattice, as we show in Ref.~\ref{fig:ZtoZZ}(b), which one can easily see is equivalent to $\overline{Z}_C \overline{Z}_T$ as depicted in Fig.~\ref{fig:OneTwistAndOneHoleQubit}.

\begin{figure}
\includegraphics{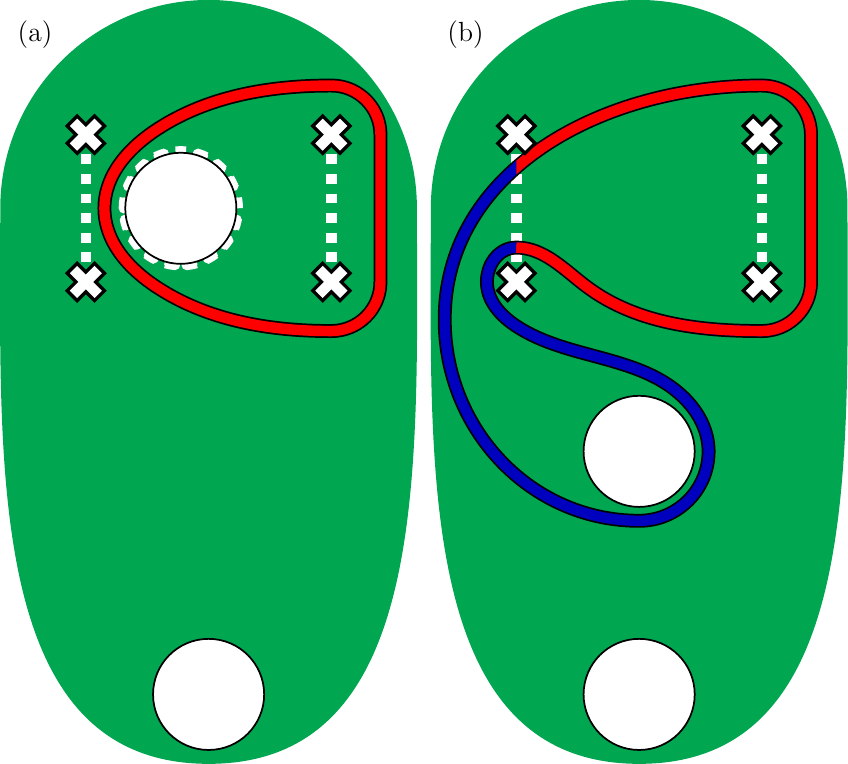}
\caption{\label{fig:XtoXX} The transformation $\overline{X}_C$ onto $\overline{X}_C \overline{X}_T$ under the code deformation scheme. (a) shows how $\overline{X}_C$ is deformed around the hole when the hole lies in between the four twists. (b) Once the hole has been deformed around the two lower twists of the lattice, and is returned to its initial position, the logical operator $\overline{X}_C$ is deformed onto a string operator that is equivalent up to stabilizer multiplication to the logical operator $\overline{X}_C \overline{X}_T$, as we show in Fig.~\ref{fig:OneTwistAndOneHoleQubit}.}
\end{figure}

We next consider the evolution of $\overline{X}_C$ as the hole is moved around the lattice. The operator $\overline{X}_C$ is a string of Pauli-Z operators that wrap around the two twists shown to the right of Fig.~\ref{fig:OneTwistAndOneHoleQubit}. While the hole has a rough boundary, it can terminate Pauli-Z strings, and as such passes transparently through the logical operator. However, once the puncture passes over the defect line, the boundary changes from rough to smooth, and as such the Pauli-Z string can no longer terminate on the hole, and must thus deform around the hole, as we see in Fig.~\ref{fig:XtoXX}(a). We continue to move the hole around its prescribed trajectory to its initial position, deforming the logical operator further, such that we finally obtain the logical operator shown in Fig.~\ref{fig:XtoXX}(b). It is easily checked that this can be deformed by stabilizer manipulations to the logical operator $\overline{X}_C \overline{X}_T$ as drawn in Fig.~\ref{fig:OneTwistAndOneHoleQubit}.

x
Finally, it is readily checked that the other two trivial relations, namely that $\overline{Z}_C \rightarrow \overline{Z}_C$, and $\overline{X}_T \rightarrow \overline{X}_T$, also hold under the code deformation scheme. With this, we show that the proposed braid completes a controlled-not gate between the two qubits shown in Fig.~\ref{fig:OneTwistAndOneHoleQubit}. We remark that, provided the holes maintain a width of $\mathcal{O}(d/4)$ and a separation $\sim d$, and the twists are mutually separated by distance $d$, then this operation is completed without reducing the distance of the code below $d$. One can also show that we achieve a controlled-phase gate between the hole-pair qubit and the quadruple-twist qubit by deforming the hole around the two twists at the left side of the diagram. We leave this as an exercise for the reader. 

Note that, once again, the effect of the braiding can be interpreted in terms of the anyons of $D(\mathbb{Z}_2)$. Given that a hole can support an electric excitation, and since two twists correspond to a single fermionic mode, which may fuse either to vacuum or to a $\psi$ excitation, the twists can be ignored and the braiding around the mode alone can be considered. The only instance in which a non-trivial braiding occurs is when both the hole and the fermionic mode are occupied by a non-trivial excitation. In this case the $e$ anyon braids around the $\psi$ excitation, which yields a $-1$ phase, and hence we observe the application of a controlled operation.

As an aside, it may be interesting to reinterpret the proposed entangling operation by code deformation, followed by a logical measurement to teleport logical information between logical qubits as a gauge-fixing operation~\cite{Paetznick13, Anderson14, Bombin15}. Specifically, this teleportation fault-tolerantly moves logical information between two different schemes of encoding. If we can regard the two logical qubits shown in Fig.~\ref{fig:OneTwistAndOneHoleQubit} as one logical qubit, and a second gauge qubit, then suitable preparation, or gauge-fixing, of the gauge qubit, followed by the entangling operation that is achieved by code deformation, and subsequent measurement of the logical qubit will teleport information from the logical qubit to the gauge qubit. After the operation is completed, the qubit that originally supported the logical information is now a gauge qubit, and the gauge qubit maintains the logical information, and as such, logical information has been switched between two different codes.

To understand this code-deformation operation as some special case of gauge fixing, we can interpret the system as the hole is braided as an intermediate code, or even a series of intermediate codes, where the nontrivial measurements that deform the code are elements of the gauge group of a subsystem code~\cite{Kribs05, Poulin05, Kribs06}. As before, the measurements we make to deform the hole can be regarded as a series of gauge-fixing operations. Finally, the measurement that is made to teleport logical information between the two codes after the entangling operation can be thought of as a third gauge fixing operation that completes the transfer of logical information. It maybe be interesting code consider this example of code switching to help us shed light on the limitations, and the potential applications of gauge fixing. We discuss this in more detail in Subsec.~\ref{Subsctn:HybridQubits}.

\subsection{Entangling twist qubits}
\label{Subsctn:EntanglingGate}
Having shown that we can perform a controlled-not operation between a quadruple-twist encoded qubit, and a hole-pair qubit, we can now use a hole-pair qubit as an ancilla qubit to perform a parity measurement between two quadruple-twist qubits. Specifically, we take the ancilla qubit and perform entangling operations between both of the qubits involved in the parity measurement, and subsequently measure the ancilla qubit.

As we have already seen, we entangle the ancilla qubit to a logical twist-quadruple qubit by braiding the hole around two of the twists that encode the qubit. It follows from this that, to perform a parity measurement between two logical qubits encoded with a quadruple of twists, we must braid a hole of an encoded ancillary hole pair around two of the twists of each qubit, and then measure the hole-pair qubit by moving the two holes together. Given the ability to perform parity measurements, we can perform a controlled-not operation between qubits encoded with quadruples of twists by measurement given an additional ancillary twist-quadruple qubit. We describe this procedure below.

\begin{figure}
\includegraphics{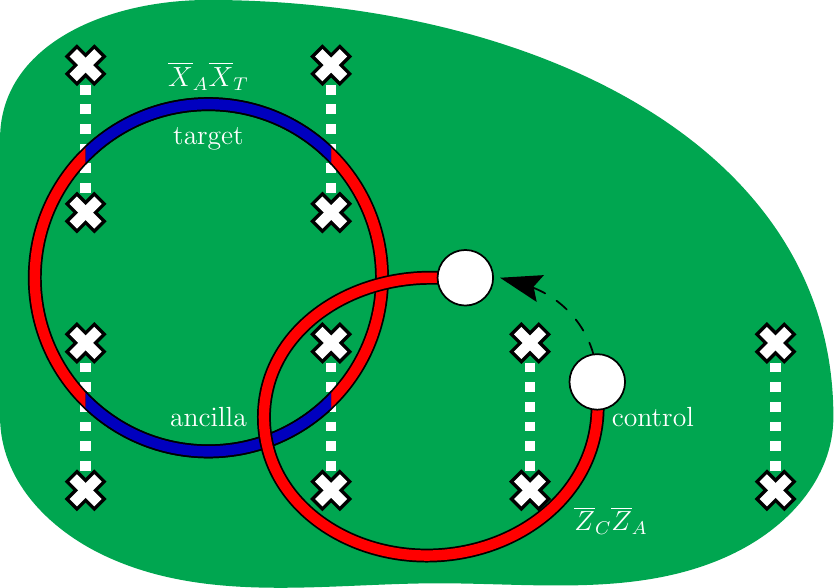}
\caption{\label{fig:TwistCNOT} Executing the two qubit parity measurements we need to perform a controlled-not gate. Qubits are encoded with quadruples of twists, where the different logical qubits, the control, target, and ancilla, are labeled in the centre of each twist quadruple. We measure $\overline{Z}_C \overline{Z}_A$ by braiding a single hole of a hole pair qubit around two twists of the control qubit and two twists of the ancilla qubit, as shown in the Figure. Upon completing this deformation, measuring the logical data encoded in the hole pair reveals the outcome of the parity measurement. We also show the $\overline{X}_A \overline{X}_T$ operator we must measure. This is again achieved by braiding one hole of a hole-pair which is prepared in $\overline{X}$ around the loop followed by the $ \overline{X}_A \overline{X}_T$ operator. Measuring $\overline{Z}_A$ completes the controlled-not gate, up to a Pauli correction.}
\end{figure}

In Fig.~\ref{fig:TwistCNOT} we show how to perform a controlled-not gate between twist qubits using parity measurements.~\cite{Terhal15}. Having prepared the ancilla qubit in the $+1$ eigenstate of $\overline{X}_A$, we then measure $\overline{Z}_C \overline{Z}_A$, followed by $\overline{X}_A \overline{X}_T$ before measuring the ancilla qubit in the computational basis, thus projecting it onto an eigenstate of $\overline{Z}_A$. 

Each of the three qubits in Fig.~\ref{fig:TwistCNOT}; the control, the ancilla and the target qubit are labeled in the centre of each twist quadruple qubit. We use an encoding where the logical operator $\overline{Z}_\alpha$($\overline{X}_\alpha$) is a closed loop enclosing two vertically(horizontally) separated twists of each quadruple, as Fig.~\ref{Fig:CornersAreTwists}(a) for $\alpha = C,\, A,\, T$. We remark also that with the chosen encoding, the total anyonic charge of a given quadruple qubit is vacuum. It follows from this that the stabilizer group contains a closed loop of Pauli-X and Pauli-Z operators that encloses all four twists of each quadruple. A consequence of this is that it does not matter which pair of vertically(horizontally) aligned twists the $\overline{Z}_\alpha$($\overline{X}_\alpha$) enclose; both are equivalent up to multiplication by stabilizer group elements, and thus have the same action on the code space.

Given these facts, together with the discussion earlier in this Section where we show that we can entangle a hole-pair qubit to quadruple-twist qubit, we can easily show how to perform a fault-tolerant controlled-not gate by parity measurements. Firstly, to perform $\overline{Z}_C \overline{Z}_A$ we prepare a pair of holes in the $+1$ eigenvalue eigenstate of logical operator $\overline{X}_h$, where we have used a lower-case index to indicate this is the logical qubit encoded by the hole pair. Next, knowing that braiding one of the two holes of the hole pair around two of the vertically aligned twists of a quadruple qubit performs a controlled-phase gate, it follows that braiding a hole around two of the vertically aligned qubits of the ancilla qubit and then two of the vertically aligned qubits of the control qubit, and then returning the hole to its initial position will perform a controlled-phase gate between the hole-pair qubit and the ancilla qubit, and a controlled-phase gate between the hole-pair qubit and the logical control qubit. In the bottom right-hand side of Fig.~\ref{fig:TwistCNOT} we show how the $\overline{X}_h$ logical operator is deformed into a logical operator which is equivalent up to stabilizers to the logical operator $\overline{Z}_C\overline{Z}_A\overline{X}_h$. Finally, following this entangling operation, measuring $\overline{X}_h$ returns the value of the fault-tolerant non-destructive parity measurement. This measurement is completed by moving the two holes of the hole pair back together, as we indicate by the black-dashed arrow in Fig.~\ref{fig:TwistCNOT}.

We must also perform a $\overline{X}_A \overline{X}_T$ gate to execute the fault-tolerant controlled-not gate. This high-weight logical operator is also shown in Fig.~\ref{fig:TwistCNOT}, where the logical operator is a string of Pauli operators that enclose two of the horizontally aligned twists of the ancilla qubit and two of the horizontally aligned qubits of the target qubit. To measure this logical operator fault-tolerantly, once again, we prepare an additional logical qubit using a pair of holes close to both the ancilla qubit and the target qubit in the $+1$ eigenstate of $\overline{X}_h$. We then deform one of the holes of the pair around two of the horizontally aligned qubits of both the ancilla qubit and the target qubit. This deformation effectively performs a controlled-not gate between the hole-pair qubit and the ancilla qubit, and a controlled-not gate between the hole-pair qubit and the target qubit. Once again, measuring the hole-pair qubit in the basis of eigenstates of the $\overline{X}_h$ operator completes the $\overline{X}_A \overline{X}_T$ parity measurement. 

Finally, we perform the fault-tolerant $\overline{Z}_A$ measurement by producing a hole-pair qubit in an eigenstate of $\overline{X}_h$, performing a controlled-phase gate as we have already described, and then measuring $\overline{X}_h$. The outcome of this measurement determines the Pauli correction we must apply to complete the measurement-only controlled-not gate. Together with the methods we have outlined in Subsec.~\ref{Subsctn:SingleQubitGates}, we can generate all of the gates of the Clifford group. Alongside noisy processes such as magic state distillation~\cite{BravyiKitaev05} we can use the Clifford group to perform universal quantum computation using our scheme.

\section{Lattice surgery within the twist framework}
\label{Sctn:Surgery}

\begin{figure}
\includegraphics{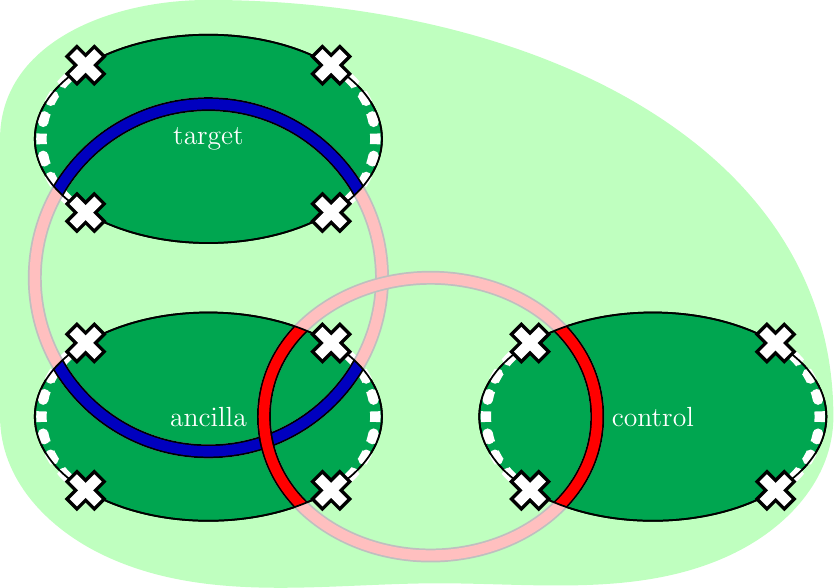}
\caption{Using the picture of twists, we can view lattice surgery as a measurement-only approach to performing logical gates between non-Abelian point particles. In bold colours we show three planar codes, and the $\overline{Z}_C \overline{Z}_A$ and $\overline{X}_A\overline{X}_T$ measurements we need to perform lattice surgery in red and blue respectively. We also show in pale colours the qubits used in the measurement-only scheme above that are not required in the lattice surgery scheme. From this perspective, we see that lattice surgery is a measurement-only topological quantum computation scheme with a significant reduction in resource costs, as we are able to perform the required parity measurements with only $\mathcal{O}(L)$ ancillary physical qubits, which is in contrast to $\mathcal{O}(L^2)$ qubits if we maintain all of the twist defects on the same lattice.  \label{fig:LatticeSurgery}}
\end{figure}

Finally, with the observation that the corners of the planar code can be regarded as Majorana modes, it is interesting to recognise that lattice surgery~\cite{Horsman12, Landahl14} is reminiscent of the measurement-only entangling gate scheme that we discuss in Subsec.~\ref{Subsctn:EntanglingGate}. 

Lattice surgery provides a way of entangling pairs of planar codes within a two-dimensional architecture via fault-tolerant logical parity measurements. This entanglement scheme is particularly interesting from a practical perspective because, in principle, the planar codes of this computational architecture can be kept well separated while they are not interacting with other qubits. In this sense, the lattice-surgery computational scheme is modular. In contrast, braiding schemes, for instance, where all the qubits are encoded over twist defects or punctures, need to be kept on a common manifold, and as such it will be necessary to design large surface code architectures with these schemes which can support all of the logical qubits that are needed to complete a computation.

In Fig.~\ref{fig:LatticeSurgery} we show how an entangling gate is performed between logical qubits encoded with planar codes via lattice surgery. The figure depicts three planar codes, an ancilla qubit, a control qubit and a target qubit, together with the $\overline{Z}_C\overline{Z}_A$ operator and the $\overline{X}_A \overline{X}_T$ operator shown in red and blue, respectively, that are used to perform entangling gates via lattice surgery~\cite{Horsman12}. The Figure also shows the additional qubits of the lattice used in the original measurement-only topological quantum-computation scheme that we have discussed above. The qubits used in the scheme above that are not required in lattice surgery are shown in pale colours. Instead, lattice surgery uses only $\mathcal{O}(L)$ qubits to perform two-qubit parity measurements. This is in contrast to the measurement-only scheme discussed above where $\mathcal{O}(L^2)$ qubits are required in between logical qubits encoded with twist quadruples on the lattice in order to maintain the distance of the code. Details on performing fault-tolerant logical parity measurements by lattice surgery are given in Refs.~\cite{Horsman12, Landahl14}.

Following this observation, it may also be interesting to explore this picture further to discover new fault-tolerant schemes for quantum computation with low resource demands using other more exotic topological models~\cite{Levin05, Koenig10, BarkeshliJianQi13}. For instance, one might also consider reinterpreting lattice surgery with the color code from the point of view of twist defects. Lattice surgery and twist defects have been considered for computation with the color code in Refs.~~\cite{Landahl14} and~\cite{Bombin11}, respectively.

\section{A hybrid encoding scheme}
\label{Sctn:Hybrids}

\begin{figure}
\includegraphics{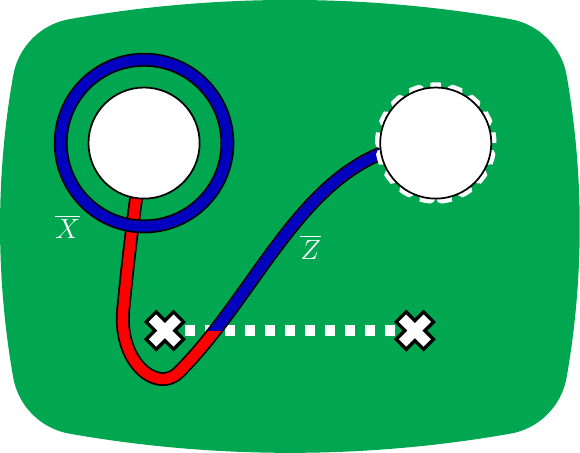}
\caption{A hybrid qubit is encoded using a pair of holes, one with a rough boundary and one with a smooth boundary, and one pair of twist defects. The two anti-commuting logical operators of the hybrid qubit are shown.  \label{Fig:HybridQubit}}
\end{figure}

Having considered several different fault-tolerant schemes for encoding and manipulating quantum information, we finally introduce a new method of encoding logical qubits that makes use of both punctures and twist defects. We call qubits encoded in this fashion hybrid qubits. Entangling operations can be achieved without additional ancilla qubits, and we can perform one non-trivial single-qubit element of the Clifford group. With these encoding we need only braid holes with different boundary types. More details on hybrid qubits are found in Ref.~\cite{Laubscher16}.

\subsection{A hybrid qubit}

We show a hybrid qubit in Fig.~\ref{Fig:HybridQubit}, together with its logical operators. The Figure shows two twist defects at the terminal points of a single defect line, together with one hole with a rough boundary, and a second hole with a smooth boundary. The logical operator $\overline{Z}$ extends from one hole, around a single twist defect and terminates at the other hole. The logical operator must follow this trajectory because the string must cross the defect line such that it can terminate at both of the punctures of the qubit. The logical operator $\overline{X}$ is a string that forms a loop which encloses a single puncture of the hybrid qubit, as seen in the Figure. To achieve code distance $d$, the two holes must each have a circumference $\mathcal{O}(d)$, the two twist defects must be separated from each other by distance $\sim d$, and the both defects must be separated from the two holes by a distance $\mathcal{O}(d/2)$.

We entangle pairs of hybrid qubits by exchanging the smooth hole of one hybrid qubit with the rough hole of the second hybrid qubit. We can also perform the single-qubit Clifford operation $B_1$ by braiding the two holes of a hybrid qubit, or by exchanging the two twist defects. It is interesting that we can combine punctures and dislocation lines to encode qubits. Considering such combined could potentially be used to discover better encoding rates of logical qubits to physical qubits than those currently known. We leave such a calculation to future work. Similar work in this direction is given in Ref.~\cite{Delfosse16, Yoder16}. However, finding the optimal rate for encoding qubits remains an open problem. In what follows we describe how to fault-tolerantly convert between different encodings that may allow us to exploit the benefits of each qubit type.

\subsection{Switching between different encodings}
\label{Subsctn:HybridQubits}

We have now identified and discussed three distinct methods of encoding logical qubits in the surface code, namely by hole pairs, twist quadruples, and with hybrid qubits; all of which have distinct capabilities of performing logical operations. Given their complementary properties, it is interesting to see that we can fault-tolerantly switch between these different encodings. We have already outlined how we can switch between a hole-pair qubit and a twist-quadruple qubit in Subsec.~\ref{Subsctn:EnganglingTwistsAndHoles}. In what follows we briefly describe how to switch between a hybrid encoding, and the other two encodings, thus providing a direct path to switch between any two of the three encodings. We remark that the switching procedures we give can be understood naturally in the anyonic picture, as they are simply transferring the anyons from one occupational mode to another without allowing them to be measured. We also point out that this idea of code switching is reminiscent of the ideas presented in Ref.~\cite{Levaillant15}, where code switching is used to complete a universal gate set with parity measurements for a particular non-Abelian anyon model.

We first consider transferring a logical qubit from a hybrid qubit to a twist qubit. To do so, we first prepare a second pair of twist defects some distance at least $\mathcal{O}(d/2)$ from the holes of the hybrid qubit, and a distance at least $\mathcal{O}(d)$ from the two twist defects of the hybrid qubit. Then, one of the two holes, say the hole with a smooth boundary of the hybrid qubit is braided around one of the two new twist defects by code deformation, and then returned to its initial position. The braid operation moves the hole across a defect line and thus changes the boundary type of the hole.

After completing the braid, we can measure a string of Pauli-Z operators that terminates at the boundaries of the two distinct holes, which maps the logical qubit onto a logical encoding of the four twists that now remain on the lattice, up to some Pauli correction. The Pauli correction we must apply is determined by the outcome of the string operator measurement. The string-operator measurement can be performed fault-tolerantly by moving the two holes together to form a single hole. The single remaining hole can subsequently be closed, leaving a single logical qubit encoded over four twists on the lattice, thus completing the code switching operation.

We can also map from a twist-quadruple qubit to onto a hybrid qubit. This is achieved by preparing a pair of holes on the lattice a distance $\mathcal{O}(d/2)$ away from the twist defects. Then, we braid one of the two holes around one of the twist defects of the twist-quadruple qubit and return the hole to its original position. Finally we measure a loop operator that encloses two twists, including the one twist that was braided with the hole, to teleport the encoded logical information onto a hybrid qubit up to a Pauli correction which is determined by the outcome of the loop measurement. Remaining on the lattice is a hybrid qubit which is made up of two holes on the lattice, and two twist defects that were not enclosed by the loop operator measurement. The other two twists that remain on the lattice can be removed by code deformation.

Finally, to map between a hybrid qubit and a hole-pair qubit, we simply braid the hole with a smooth boundary of the hybrid qubit around one of the two twist defects of the hybrid qubit which transforms the boundary type of the braided hole. We are then free to remove the twists from the lattice as the logical information is now preserved in the hole-pair qubit. The reverse operation can be performed such that a hole-pair qubit is mapped onto a hybrid qubit by preparing pair of twist defects on the lattice, and then braiding one of the holes of the hole pair around one of the new twist defects. The two twist defects, and the two holes now compose a hybrid qubit describing the logical information that was initially encoded by the hole-pair qubit.

Given that the three different encodings of logical qubits on the surface code all have complementary properties, it is interesting to find the most resource efficient method of encoding and manipulating logical qubits. To remind the reader, the twist qubit can generate all of the single-qubit Clifford operations by braiding, but requires a logical ancilla qubit to perform entangling gates using parity measurements, the double code of Subsec.~\ref{subsec:doublediamond} being a notable exception. In contrast, hole-pair qubits can be entangled readily, but do not achieve single-qubit Clifford gates by braiding. Hybrid qubits fall in the middle ground of these two examples, as they can be directly entangled without ancilla, and can perform a subset of single-qubit Clifford rotations. Given that we are also capable of efficiently switching between these different types of qubits, it may be interesting to try to discover more efficient computational schemes in space-time resource costs using code switching. We also mention again that it may be an interesting direction of study to reinterpret the examples of code deformation gauge-fixing scheme from a more fundamental point of view.

\section{Concluding remarks}
\label{Sctn:Conclusion}

To summarise, we have unified several methods of manipulating logical qubits with the surface code. Notably, we have demonstrated new code deformation schemes to implement the full Clifford group using a small number of weight-five local measurements over the planar code lattice. In contrast, surface code quantum computation using only hole defects or standard lattice surgery may be more resource intensive once the implementation of full algorithms are considered, as with these architectures we must prepare and maintain additional logical qubits in eigenstates of the Pauli-Y matrix via a distillation scheme to complete the Clifford group. Alternatively, the two-dimensional color code, which achieves the full Clifford group transversally, requires weight-six stabilizer measurements, which we expect to be more challenging to perform in the laboratory. With these considerations, we argue that the new deformation procedures we have considered may lead to fault-tolerant quantum-computational schemes with lower resource costs than previously considered architectures. To interrogate our corner braiding scheme further, we should examine how it behaves at the circuit level under a realistic noise model as gates are performed. We leave such an analysis to future work.

We have also built on the analogy between the corners of the planar code, twists and Ising anyons to show that lattice surgery fits into the more conventional picture of measurement-only topological quantum computation. We suggest that this observation may be extended to other topological phases with boundaries to develop other fault-tolerant quantum computational schemes with lattice surgery. Certainly, it may be instructive to find modular quantum computational models where a universal gate set is achieved between qubits encoded on some suitably chosen topological substrates via fault-tolerant logical parity measurements. It might also be interesting to adapt the schemes we develop here for use in a fault-tolerant measurement-based scheme~\cite{Raussendorf07, Raussendorf07a, Brell15, Bolt16} for quantum computation. Such an extension may make some of the present ideas experimentally amenable to a linear optical architecture~\cite{Rudolph16}.

\begin{acknowledgements}
The authors acknowledge J. Auger, H. Bomb\'{i}n, M. Kastoryano, D. Loss, J. Pachos and T. Stace for helpful and encouraging conversations, and we thank C. Brell for discussions on adapting our proposal to a fault-tolerant measurement-based scheme. Furthermore, we are particularly grateful to B. Terhal for an enlightening discussions on code deformation. We also extend our gratitude to S. Bravyi and N. Delfosse for comments on earlier drafts of the manuscript which have helped us improve the clarity of our presentation.  BJB is supported by the Villum Foundation. JRW is supported by the NCCR QSIT. MSK is supported by CRC183, and acknowledges the hospitality of the University of Leeds, where part of this work was completed.
\end{acknowledgements}

\bibstyle{plain}

\end{document}